\documentclass[12pt,a4size]{article}
\usepackage{mathtext}        % ???? ?y??? py????? ?y??? ? ??p?y??
\usepackage{amsmath}
\usepackage{textcomp}
\usepackage{graphicx}
\usepackage{indentfirst}
\begin{document}

\begin{center}
{\bf The Hubble flow: an observer's perspective}\\

Alexey V. Toporensky, Sergei B. Popov\\
(Lomonosov Moscow State University, Sternberg Astronomical Institute)

  \end{center}

\vskip 0.1cm

\begin{abstract}
%{\bf Abstract}
%  \end{center}

In this methodological note we discuss some details and peculiarities of the cosmic expansion as viewed
by a realistic observer. We show that the velocity $v_\Theta$ related to a
change (measured by observer's clock) of the angular distance, plays an
important role in formation of a meaningful observed picture of the expansion
of the universe.
Usage of this velocity and the angular distance (in addition to the
standard approach --- proper distance and corresponding velocity) 
allows to present the cosmic expansion in a more
illustrative manner. These parameters play a key role in
visualization of the expansion.

\end{abstract}

% э┘ ╥┴╙╙═┴╘╥╔╫┴┼═ ╬┼╦╧╘╧╥┘┼ ╧╙╧┬┼╬╬╧╙╘╔ ╦╧╙═╧╠╧╟╔▐┼╙╦╧╟╧ ╥┴╙█╔╥┼╬╔╤ ╙ ╘╧▐╦╔
%┌╥┼╬╔╤ ╬┴┬╠└─┴╘┼╠╤. ї╦┴┌┘╫┴┼╘╙╤ ╔ ╧┬╙╒╓─┴┼╘╙╤ ╧╙╧┬┴╤ ╥╧╠╪ ╙╦╧╥╧╙╘╔ ╔┌═┼╬┼╬╔╤
%╒╟╧╠╧╫╧╟╧ ╥┴╙╙╘╧╤╬╔╤, ╔┌═┼╥╤┼═╧╩ ╨╧ ▐┴╙┴═ ╬┴┬╠└─┴╘┼╠╤, --- $v_\Theta$, --- ╫
%╙╧┌─┴╬╔╔ ┴─┼╦╫┴╘╬╧╟╧ ╧┬╥┴┌┴ ╨╧╘┼╬├╔┴╠╪╬╧ ╬┴┬╠└─┴┼═╧╩ ╦┴╥╘╔╬┘ ╥┴╙█╔╥┼╬╔╤.
%щ╙╨╧╠╪┌╧╫┴╬╔┼ ▄╘╧╩ ╙╦╧╥╧╙╘╔ ╔ ╒╟╠╧╫╧╟╧ ╥┴╙╙╘╧╤╬╔╤ (╫ ─╧╨╧╠╬┼╬╔┼ ╦
%╘╥┴─╔├╔╧╬╬╧═╒ ╙╨╧╙╧┬╒ ╔╠╠└╙╘╥┴├╔╔) ╨╧┌╫╧╠╤┼╘
%┬╧╠┼┼ ╬┴╟╠╤─╬╧ ╨╥┼─╙╘┴╫╔╘╪ ╦┴╥╘╔╬╒ ╦╧╙═╧╠╧╟╔▐┼╙╦╧╟╧ ╥┴╙█╔╥┼╬╔╤  
%╔ ─╧╠╓╬╧ ╤╫╠╤╘╪╙╤ ╦╠└▐┼╫┘═ ╨╥╔ ┼╟╧ ┴─┼╦┴╘╬╧╩ ╫╔┌╒┴╠╔┌┴├╔╔.   

%\vskip 1cm

%\newpage

\section{Introduction}

 How do we imagine the cosmic expansion?
Usually this is a traditional image used in popular science, as well as
in textbooks and even monographs. This is a ``bird's-eye view'' or a ``god's
view'', when we find ourselves out of our space observing it from outside. 
For example, we imagine an inflating ball, or a stretching surface, which
represent our expanding universe. More than that, it is convenient to
imagine all points on this ball (or the surface) visible similtaniously,
i.e. we see the whole picture ``as it is now''.  Hence, we not only observe
the universe ``from outside'', but also ``see'' all its points
at the same time.

%ы┴╦ ═┘ ╨╥┼─╙╘┴╫╠╤┼═ ╙┼┬┼ ╥┴╙█╔╥┼╬╔┼ ╫╙┼╠┼╬╬╧╩?
%я┬┘▐╬╧ ╨┼╥┼─ ╟╠┴┌┴═╔ ╫╧┌╬╔╦┴┼╘ ╧┬╥┴┌, ╘╥┴─╔├╔╧╬╬╧ ╞╔╟╒╥╔╥╒└▌╔╩ ╦┴╦ ╫
%╬┴╒▐╬╧-╨╧╨╒╠╤╥╬╧╩, ╘┴╦ ╔ ╫ ╒▐┼┬╬╧╩, ╔ ─┴╓┼ ╫ ╙╨┼├╔┴╠╪╬╧╩ ╠╔╘┼╥┴╘╒╥┼.  
%№╘╧ ╫┌╟╠╤─ ``╙ ╫┘╙╧╘┘ ╨╘╔▐╪┼╟╧ ╨╧╠┼╘┴'', ╔╠╔ ``╘╧▐╦┴ ┌╥┼╬╔╤ ┬╧╟┴'', 
%╦╧╟─┴ ═┘ ╨╥┼─╙╘┴╫╠╤┼═ ╙┼┬╤ ╫╬┼ ╬┴█┼╟╧ ╨╥╧╙╘╥┴╬╙╘╫┴, ╬┴┬╠└─┴╤ ┼╟╧ ╦┴╦ ┬┘
%╙╧ ╙╘╧╥╧╬┘.  
%ю┴╨╥╔═┼╥, ═┘ ╨╥┼─╙╘┴╫╠╤┼═ ╨┼╥┼─ ╙╧┬╧╩ ╥┴┌─╒╫┴└▌╔╩╙╤ █┴╥╔╦ ╔╠╔
%╥┴╙╘╤╟╔╫┴└▌╒└╙╤ ╨╠╧╙╦╧╙╘╪, ╦╧╘╧╥┘┼ ╔ ╧╠╔├┼╘╫╧╥╤└╘ ╬┴█╒ ╥┴╙█╔╥╤└▌╒└╙╤
%╫╙┼╠┼╬╬╒└.
%т╧╠┼┼ ╘╧╟╧, ╬┴═ ╒─╧┬╬╧ ╨╥┼─╙╘┴╫╠╤╘╪ ╫╙┼ ╘╧▐╦╔ ▄╘╧╟╧ █┴╥╔╦┴ ╔╠╔ ╨╠╧╙╦╧╙╘╔ ╦┴╦
%┬┘ ╬┴┬╠└─┴┼═┘═╔  ╧─╬╧╫╥┼═┼╬╬╧, ╘.┼.  ═┘ ╫ ╥┴═╦┴╚ 
%╘┴╦╧╩ ╔╠╠└╙╘╥┴├╔╔ ╫╧╙╨╥╔╬╔═┴┼═ ╫╙└ ╦┴╥╘╔╬╒ ``╦┴╦ ╧╬┴ ┼╙╘╪ ╙┼╩▐┴╙''.
%Ї┴╦╔═ ╧┬╥┴┌╧═, ═┘ ╧┬┘▐╬╧ ╬┼ ╘╧╠╪╦╧ ╨╥┼─╙╘┴╫╠╤┼═ ╙┼┬┼ ╫╙┼╠┼╬╬╒└ ``╔┌╫╬┼'', ╬╧ ╔
%``╫╔─╔═'' ╫╙┼ ┼┼ ╘╧▐╦╔ ╫ ╧─╔╬ ╔ ╘╧╘ ╓┼ ═╧═┼╬╘ ╫╥┼═┼╬╔.

 This is a useful image (maybe, it is necessary for understanding), however,
a real observer never sees such a picture, it is impossible in principle. How
the expansion would look like for an ``internal'' observer?

% №╘╧ ╚╧╥╧█╔╩ ╧┬╥┴┌ (╔, ╫╧┌═╧╓╬╧, ╬┼╧┬╚╧─╔═┘╩ ─╠╤ ╨╧╬╔═┴╬╔╤), ╬╧ ╙╧┬╙╘╫┼╬╬╧ ╥┼┴╠╪╬┘╩
%╬┴┬╠└─┴╘┼╠╪ ╘┴╦╒└ ╦┴╥╘╔╬╒ ╬╔╦╧╟─┴ ╬┼ ╙═╧╓┼╘ ╒╫╔─┼╘╪, ─┴╓┼ ╘┼╧╥┼╘╔▐┼╙╦╔. 
%ы┴╦ ╓┼ ┬╒─┼╘ ╫┘╟╠╤─┼╘╪ ╥┴╙█╔╥┼╬╔┼ ╫╙┼╠┼╬╬╧╩ ╙ ╘╧▐╦╔ ┌╥┼╬╔╤ ╥┼┴╠╪╬╧╟╧ ╬┴┬╠└─┴╘┼╠╤ ``╫╬╒╘╥╔''?

 Let us imagine that we can make observations with arbitrarily high
precision, or that we can observe long enough to measure with existing
instruments changes in parameters of distant sources due to their recession. 
How can we better illustrate the results presenting the cosmic expansion?
Or, let our goal be a realistic 3D model (for example, for a planetarium) of
the picture that an observer sees in an expanding universe. Which parameters
fit better for such a task? In particular, which velocity we have to choose
to illustrate recession of galaxies?

%Ё╥┼─╙╘┴╫╔═ ╙┼┬┼, ▐╘╧ ═┘ ═╧╓┼═ ╨╥╧╫╧─╔╘╪ ╬┴┬╠└─┼╬╔╤ ╙╧ ╙╦╧╠╪ ╒╟╧─╬╧ ╫┘╙╧╦╧╩ ╘╧▐╬╧╙╘╪└, ╔╠╔ ╓┼ ═╧╓┼═
%╬┴┬╠└─┴╘╪ ─╧╙╘┴╘╧▐╬╧ ─╧╠╟╧, ▐╘╧┬┘ ╔┌═┼╥╔╘╪ ╙ ╨╧═╧▌╪└ ╙╒▌┼╙╘╫╒└▌╔╚ ╔╬╙╘╥╒═┼╬╘╧╫ ╔┌═┼╬┼╬╔┼ 
%╚┴╥┴╦╘┼╥╔╙╘╔╦ ─┴╠┼╦╔╚ ╧┬▀┼╦╘╧╫ ╔┌-┌┴ ╔╚ ╒─┴╠┼╬╔╤.  ы┴╦ ╬┴═ ╠╒▐█┼, ╬┴╟╠╤─╬┼┼ ╨╥┼─╙╘┴╫╔╘╪
%╥┼┌╒╠╪╘┴╘┘, ╬┼╨╧╙╥┼─╙╘╫┼╬╬╧ ╧╘╥┴╓┴└▌╔┼ ╦╧╙═╧╠╧╟╔▐┼╙╦╧┼ ╥┴╙█╔╥┼╬╔┼?
%щ╠╔ ╨╒╙╘╪ ╬┴█┼╩ ┌┴─┴▐┼╩ ┬╒─┼╘ ╙╧┌─┴╬╔┼ ╥┼┴╠╔╙╘╔▐╬╧╟╧ 3D ═╧─┼╠╔╥╧╫┴╬╔╤ 
%(╙╦┴╓┼═, ─╠╤ ╨╠┴╬┼╘┴╥╔╤) ╘╧╩ ╦┴╥╘╔╬┘, ╦╧╘╧╥╒└ ╒╫╔─╔╘ ╬┴┬╠└─┴╘┼╠╪ ╫ ╥┴╙█╔╥╤└▌┼╩╙╤ ╫╙┼╠┼╬╬╧╩. 
%ў┘┬╧╥ ╦┴╦╔╚ ╨┴╥┴═┼╘╥╧╫ ╨╥┼─╨╧▐╘╔╘┼╠╪╬┼┼ ╙ ╘╧▐╦╔ ┌╥┼╬╔╤ ╫┘╨╧╠╬┼╬╔╤ ╘┴╦╧╩ ┌┴─┴▐╔? 
%ў ▐┴╙╘╬╧╙╘╔, ╦┴╦╒└ ╙╦╧╥╧╙╘╪ ╬┴═ ╬╒╓╬╧ ┬╒─┼╘ ╫┘┬╥┴╘╪ ─╠╤ ╧╨╔╙┴╬╔╤ ╥┴┌╠┼╘┴
%╟┴╠┴╦╘╔╦?

 In this methodological note we demonstrate that one of the best choice is
the velocity related to the so-called angular distance, $d_\Theta$.
We discuss some characteristics of this parameter, and show how it behaves
in different universes. This approach, in our opinion, is an important
supplement to the traditional illustration (``god's view''), and helps to form a more
adequate image  of an expanding universe. This is essential, as many
phenomena in cosmology are not vivid and, at first glance, contradict
common sense (which includes a very advanced ``common sense'', see for
example an interesting discussion about feasibility of superluminal velocities
of the Hubble flow and different corresponding misconception and mistakes in
\cite{murd}, and also in \cite{davis01} and references therein).
Insufficient clarity of cosmic phenomena results, for example among students, 
in difficulties in qualitative understanding of the expansion of the
universe and associated effects. This is what we hope to overcome.

%ў ▄╘╧╩ ═┼╘╧─╔▐┼╙╦╧╩ ┌┴═┼╘╦┼ ═┘ ╨╧╦┴┌┘╫┴┼═, ▐╘╧ ╧─╬╧╩ ╔┌ ╬┴╔┬╧╠┼┼
%╔╠╠└╙╘╥┴╘╔╫╬┘╚ ╫┼╠╔▐╔╬ ┬╒─┼╘
%╙╦╧╥╧╙╘╪, ╙╫╤┌┴╬╬┴╤ ╙ ╘.╬.  ╒╟╠╧╫┘═ ╥┴╙╙╘╧╤╬╔┼═, $d_\Theta$.  
%э┘ ╧┬╙╒╓─┴┼═ ╬┼╦╧╘╧╥┘┼ ╧╙╧┬┼╬╬╧╙╘╔ ▄╘╧╩ ╫┼╠╔▐╔╬┘, ┴ ╘┴╦╓┼ ┼┼ ╨╧╫┼─┼╬╔┼
%╫╧ ╫╙┼╠┼╬╬┘╚ ╙ ╥┴┌╠╔▐╬┘═ ╬┴┬╧╥╧═ ╨┴╥┴═┼╘╥╧╫.  Ї┴╦╧╩ ╨╧─╚╧─, ╦┴╦ ╬┴═
%╨╥┼─╙╘┴╫╠╤┼╘╙╤, ╤╫╠╤╤╙╪ ╙╒▌┼╙╘╫┼╬╬┘═ ─╧╨╧╠╬┼╬╔┼═ ╘╥┴─╔├╔╧╬╬╧╩ ╔╠╠└╙╘╥┴├╔╔
%("╦┴╥╘╔╬┘ ┬╧╟┴"), ╨╧┌╫╧╠╔╘
%╙╧┌─┴╘╪ ┬╧╠┼┼ ┴─┼╦╫┴╘╬┘╩ ╧┬╥┴┌ ╥┴╙█╔╥╤└▌┼╩╙╤ ╫╙┼╠┼╬╬╧╩. №╘╧ ╫┴╓╬╧, ╘.╦. 
%═╬╧╟╔┼ ╤╫╠┼╬╔╤
%╫ ╦╧╙═╧╠╧╟╔╔ ╬┼ ╙╠╔█╦╧═ ╬┴╟╠╤─╬┘ ╔, ╬┴ ╨┼╥╫┘╩ ╫┌╟╠╤─, ╨╥╧╘╔╫╧╥┼▐┴╘ ┌─╥┴╫╧═╒
%╙═┘╙╠╒
%(╫ ╘╧═ ▐╔╙╠┼ ╔ ╫┼╙╪═┴ ``╨╥╧─╫╔╬╒╘╧═╒'' ┌─╥┴╫╧═╒ ╙═┘╙╠╒, 
%╙═., ╬┴╨╥╔═┼╥, ╔╬╘┼╥┼╙╬╒└ ─╔╙╦╒╙╙╔└ ╧ ─╧╨╒╙╘╔═╧╙╘╔
%╙╫┼╥╚╙╫┼╘╧╫┘╚ ╙╦╧╥╧╙╘┼╩ ╚┴┬┬╠╧╫╙╦╧╟╧ ╨╧╘╧╦┴ ╔ ╫╧┌╬╔╦┴└▌╔╚ ╬┼─╧╥┴┌╒═┼╬╔╤╚ ╔
%╧█╔┬╦┴╚ ╫ \cite{murd}, ┴ ╘┴╦╓┼ ╫ \cite{davis01} ╔ ├╔╘╔╥╒┼═┘╚ ╘┴═ ╥┴┬╧╘┴╚).  
%ю┼─╧╙╘┴╘╧▐╬┴╤ ╬┴╟╠╤─╬╧╙╘╪ ╫┘┌┘╫┴┼╘, ╫ ▐┴╙╘╬╧╙╘╔ ╒ ╙╘╒─┼╬╘╧╫, ┌┴╘╥╒─╬┼╬╔╤ ╫
%╦┴▐┼╙╘╫┼╬╬╧═ ╨╧╬╔═┴╬╔╔
%╦╧╙═╧╠╧╟╔▐┼╙╦╧╟╧ ╥┴╙█╔╥┼╬╔╤ ╔ ╙╫╤┌┴╬╬┘╚ ╙ ╬╔═ ▄╞╞┼╦╘╧╫.   щ═┼╬╬╧ ▄╘╧ ═┘ ╔
%╨╧╨┘╘┴┼═╙╤ ╨╥┼╧─╧╠┼╘╪.

%\newpage

\section{Distances in cosmology}

Definitions and properties of different distance measures used in cosmology
can be found in any standard textbook on cosmology
(see, for example, \cite{wein}).  In the present section we review basics of
several main definitions of cosmological distances as we extensively use
some of them below.

%Є┴╙╙═╧╘╥┼╬╔┼ ╥┴┌╠╔▐╬┘╚ ╥┴╙╙╘╧╤╬╔╩, ╫╫╧─╔═┘╚ ╫ ╦╧╙═╧╠╧╟╔╔, ╤╫╠╤┼╘╙╤
%╙╘┴╬─┴╥╘╬┘═ ▄╠┼═┼╬╘╧═ ╔┌╠╧╓┼╬╔╤ ╧╙╬╧╫ ▄╘╧╩ ╬┴╒╦╔ (╙═., ╬┴╨╥╔═┼╥, \cite{wein}). 
%ў ▄╘╧═ ╥┴┌─┼╠┼ ═┘ ╦╥┴╘╦╧ ╙╒══╔╥╒┼═ ┬┴┌╧╫┘┼ ╨╧╬╤╘╔╤, ╦┴╙┴└▌╔┼╙╤ ╥┴╙╙╘╧╤╬╔╩ ╫
%╦╧╙═╧╠╧╟╔╔, ╘.╦. ╧╬╔ ┬╒─╒╘ ╬┼╧┬╚╧─╔═┘ ╬┴═ ╫ ─┴╠╪╬┼╩█┼═.

Let us consider a Friedmann universe. In this space we can make
a special choice of time coordinate for which
all surfaces of constant time are homogenious (this defines the cosmic time).  
It is natural to use it in future considerations.  

%ў ╦┴▐┼╙╘╫┼ ╦╧╬╦╥┼╘╬╧╟╧ ╨╥╔═┼╥┴ ╦╧╙═╧╠╧╟╔▐┼╙╦╧╩ ═╧─┼╠╔ ═┘ ┬╒─┼═ ╥┴╙╙═┴╘╥╔╫┴╘╪ ╞╥╔─═┴╬╧╫╙╦╒└
%╫╙┼╠┼╬╬╒└.
%·─┼╙╪ ╙╠┼─╒┼╘ ╬┴╨╧═╬╔╘╪, ▐╘╧ ╫╧ ╞╥╔─═┴╬╧╫╙╦╧╩ ╫╙┼╠┼╬╬╧╩ ╙╒▌┼╙╘╫╒┼╘
%╫┘─┼╠┼╬╬┘╩ ╫┘┬╧╥ ╫╥┼═┼╬╬╧╩ ╦╧╧╥─╔╬┴╘┘, ╨╥╔ ╦╧╘╧╥╧═
%╙╧╧╘╫┼╘╙╘╫╒└▌╔┼ ╨╥╧╙╘╥┴╬╙╘╫┼╬╬┘┼ ╙┼▐┼╬╔╤ ╧─╬╧╥╧─╬┘, ╔ ┼╙╘┼╙╘╫┼╬╬╧
%╨╧╠╪┌╧╫┴╘╪╙╤ ╔═┼╬╬╧ ╘┴╦╔═ ``╦╧╙═╔▐┼╙╦╔═ ╫╥┼═┼╬┼═''
%(┴╬╟╠╔╩╙╦╧┼ cosmic time ╤╫╠╤┼╘╙╤ ╧┬▌┼╨╥╔╬╤╘┘═ ╘┼╥═╔╬╧═ ╔ ═┘ ─╠╤ ╒─╧┬╙╘╫┴
%┬╒─┼═ ╨╧╠╪┌╧╫┴╘╪╙╤ ╥╒╙╙╦╧╩ ╦┴╠╪╦╧╩ ╙ ╬┼╟╧).

The flat Friedmann metric has the usual form:

$$
ds^2=c^2dt^2-a(t)^2dl^2.
$$
Here $t$ is cosmic time.  
It is worth noting that cosmic time is not directly available to an ``internal'' observer
who sees the universe as inhomogenious (the farther --- the denser).
The second term in the equation represents the Hubble flow ---
distant objects recede due to increasing scale
factor $a$, while their comoving coordinates do not change.  

%э╧╓╬╧ ┌┴╨╔╙┴╘╪ ╨╠╧╙╦╒└ ═┼╘╥╔╦╒ ц╥╔─═┴╬┴:
%
%$$
%ds^2=c^2dt^2-a(t)^2dl^2.
%$$
%·─┼╙╪ $t$ --- ╦╧╙═╔▐┼╙╦╧┼ ╫╥┼═╤. ў╘╧╥╧╩ ▐╠┼╬ ╙╨╥┴╫┴ ╧╘╥┴╓┴┼╘ ╙╒▌┼╙╘╫╧╫┴╬╔┼ ╚┴┬┬╠╧╫╙╦╧╟╧ ╨╧╘╧╦┴:
%╫ ╨╥┼╬┼┬╥┼╓┼╬╔╔ ╨┼╦╒╠╤╥╬┘═╔ ╙╦╧╥╧╙╘╤═╔ ─┴╠┼╦╔┼ ╧┬▀┼╦╘┘ ╒─┴╠╤└╘╙╤ ─╥╒╟ ╧╘
%─╥╒╟┴ ┌┴ ╙▐┼╘ ╒╫┼╠╔▐┼╬╔╤ ═┴╙█╘┴┬╬╧╟╧ ╞┴╦╘╧╥┴ ╨╥╔ ╙╧╚╥┴╬┼╬╔╔ ╔╚ ╙╧╨╒╘╙╘╫╒└▌╔╚ ╦╧╧╥─╔╬┴╘ ╬┼╔┌═┼╬╬┘═╔.

As for comoving coordinates, it is natural to introduce a spherical system
with an observer at the origin.
Then distances and velocities defined below depend only on the radial
comoving coordinate $\chi$.  
We will consider universes filled with a one-component perfect
fluid, in this case all necessary formulae can 
be obtained in explicit (and rather simple) form.  Indeed, for the equation
of state we have: 

%\begin{equation}
$$
p=w \rho c^2,
$$
%\end{equation}
where $p$ is pressure, $\rho$ is matter density, and $c$ is the
speed of light.
The homogenious solution for the time evolution of the scale factor is 
$a\sim t^{1/\alpha}$, where $\alpha=3(w+1)/2$.
As the Hubble parameter $H=\dot a/a=1/(\alpha t)$ and $t\propto a^{\alpha}$, 
we can, using the definition of the redshift
$1+z(t)=a(t_0)/a(t)$, (where $a(t_0)$ is the scale factor at the present
time),
express the Hubble parameter in terms of the redshift: $H=H_0 (1+z)^\alpha$ 
($H_0$ is its present value).

%Ё╥╔ ╧╨╥┼─┼╠┼╬╔╔ ╙╧╨╒╘╙╘╫╒└▌╔╚ ╦╧╧╥─╔╬┴╘ ┼╙╘┼╙╘╫┼╬╬╧ ╫╫┼╙╘╔ ╙╞┼╥╔▐┼╙╦╒└ ╙╔╙╘┼═╒ ╙ ├┼╬╘╥╧═ ╫ ╘╧▐╦┼
%╬┴╚╧╓─┼╬╔╤ ╬┴┬╠└─┴╘┼╠╤, ╘╧╟─┴ ┼─╔╬╙╘╫┼╬╬╧╩ ╫┼╠╔▐╔╬╧╩, ╧╘ ╦╧╘╧╥╧╩ ┬╒─╒╘ ┌┴╫╔╙┼╘╪ ╧╨╥┼─┼╠╤┼═┘┼ ╬╔╓┼
%╥┴╙╙╘╧╤╬╔╤ ╔ ╙╦╧╥╧╙╘╔ ╚┴┬┬╠╧╫╙╦╧╟╧ ╨╧╘╧╦┴ ┬╒─┼╘ ╥┴─╔┴╠╪╬┴╤ ╙╧╨╒╘╙╘╫╒└▌┴╤ ╦╧╧╥─╔╬┴╘┴ $\chi$.
%э┘ ┬╒─┼═ ╥┴╙╙═┴╘╥╔╫┴╘╪ ╘╧╠╪╦╧ ╘┴╦╔┼ ═╧─┼╠╔, ╫ ╦╧╘╧╥┘╚ ╫╙┼╠┼╬╬┴╤ ┌┴╨╧╠╬┼╬┴
%┬┴╥╧╘╥╧╨╬╧╩ (╘.┼., ╘┴╦╧╩, ▐╘╧ ╨╠╧╘╬╧╙╘╪ ┌┴╫╔╙╔╘ ╠╔█╪ ╧╘
%╧─╬╧╟╧ ╨┴╥┴═┼╘╥┴ --- ─┴╫╠┼╬╔╤) ═┴╘┼╥╔┼╩ ╧─╬╧╟╧ ╫╔─┴, 
%▐╘╧ ╨╧┌╫╧╠╤┼╘ ╨╧╠╒▐╔╘╪ ╫╙┼ ╬┼╧┬╚╧─╔═┘┼ ╞╧╥═╒╠┘ ╫ ┌┴═╦╬╒╘╧═ (╔ ─╧╙╘┴╘╧▐╬╧
%╨╥╧╙╘╧═) ╫╔─┼. ф┼╩╙╘╫╔╘┼╠╪╬╧, ─╠╤ ╫╙┼╠┼╬╬╧╩, ┌┴╨╧╠╬┼╬╬╧╩ ╫┼▌┼╙╘╫╧═ ╙ ╒╥┴╫╬┼╬╔┼═ ╙╧╙╘╧╤╬╔╤
%%\begin{equation}
%$$
%p=w \rho c^2,
%$$
%%\end{equation}
%╟─┼ $p$ --- ─┴╫╠┼╬╔┼, $\rho$ --- ╨╠╧╘╬╧╙╘╪, ┴ $c$ --- ╙╦╧╥╧╙╘╪ ╙╫┼╘┴,
%╧─╬╧╥╧─╬╧┼ ╥┼█┼╬╔┼ ╫┘╥┴╓┴┼╘╙╤ ╙╘┼╨┼╬╬┘═ ┌┴╦╧╬╧═: $a\sim t^{1/\alpha}$, ╟─┼ $\alpha=3(w+1)/2$.
%Ё╧╙╦╧╠╪╦╒ ╨╧╙╘╧╤╬╬┴╤ ш┴┬┬╠┴ $H=\dot a/a=1/(\alpha t)$ ╔ $t\propto a^{\alpha}$, 
%═┘ ═╧╓┼═, ╔╙╨╧╠╪┌╒╤ ╧╨╥┼─┼╠┼╬╔┼ ╦╥┴╙╬╧╟╧ ╙═┼▌┼╬╔╤
%$1+z(t)=a(t_0)/a(t)$, ╟─┼ $a(t_0)$ --- ═┴╙█╘┴┬╬┘╩ ╞┴╦╘╧╥ ╫ ╬┴╙╘╧╤▌╔╩ ═╧═┼╬╘,
%╫┘╥┴┌╔╘╪ ╚┴┬┬╠╧╫╙╦╔╩ ╨┴╥┴═┼╘╥ ╫ ╘┼╥═╔╬┴╚ ╦╥┴╙╬╧╟╧ ╙═┼▌┼╬╔╤: $H=H_0 (1+z)^\alpha$, 
%$H_0$ --- ╙╧╫╥┼═┼╬╬┴╤ ╨╧╙╘╧╤╬╬┴╤ ш┴┬┬╠┴.

%ы╠└▐┼╫┘═ ╨┴╥┴═┼╘╥╧═ ┬╒─┼╘ ╘.╬. ╙╧╨╒╘╙╘╫╒└▌┴╤ ╦╧╧╥─╔╬┴╘┴. №╘╧ ┼╙╘┼╙╘╫┼╬╬┘╩
%╫┘┬╧╥ ╦╧╧╥─╔╬┴╘, ╙╫╤┌┴╬╬┘╩ ╙ ╬┴┬╠└─┴╘┼╠╤═╔, ╦╧╘╧╥┘┼ ╫╔─╤╘ ╫╙┼╠┼╬╬╒└
%╔┌╧╘╥╧╨╬╧╩. ш┴┬┬╠╧╫╙╦╧┼ ╥┴╙█╔╥┼╬╔┼ ╬┼ ═┼╬╤┼╘ ▄╘╒ ╦╧╧╥─╔╬┴╘╒. Ё┼╦╒╠╤╥╬┘═╔
%╙╦╧╥╧╙╘╤═╔, ╙╫╤┌┴╬╬┘═╔ ╙ ╫┌┴╔═╧─┼╩╙╘╫╔┼═ ╧┬▀┼╦╘╧╫ ─╥╒╟ ╙ ─╥╒╟╧═, ╦┴╦ ╨╥┴╫╔╠╧
%═╧╓╬╧ ╨╥┼╬┼┬╥┼▐╪, ┼╙╠╔ ═┘ ╟╧╫╧╥╔═ ╧ ┬╧╠╪█╔╚ ═┴╙█╘┴┬┴╚, ┴ ╘╧╠╪╦╧ ╘┴╦╔┼ ╬┴╙ ╔
%┬╒─╒╘ ╔╬╘┼╥┼╙╧╫┴╘╪.

Using the light propagation equation
$ds^2=0$, the comoving coordinate of the object observable now at 
some redshift $z$ can be obtained 
in a standard way for a given equation of state parameter $\alpha$:

$$
\chi=\frac{c}{a(t_0) H_0}\int_0^z \frac{dz}{H(z)} =\frac{c}{a(t_0) H_0} \frac{1}{1-\alpha}
[(1+z)^{1-\alpha}-1].
$$ 
%щ╙╨╧╠╪┌╒╤ ┌┴╦╧╬ ╥┴╙╨╥╧╙╘╥┴╬┼╬╔╤ ╙╫┼╘┴ $ds^2=0$, ╙╘┴╬─┴╥╘╬┘═ ╧┬╥┴┌╧═
% ╨╧╠╒▐┴┼═ ╧┬▌┼┼ ╫┘╥┴╓┼╬╔┼ ─╠╤ ╙╧╨╒╘╙╘╫╒└▌┼╩ ╦╧╧╥─╔╬┴╘┘ ╧┬▀┼╦╘┴, ╬┴┬╠└─┴┼═╧╟╧
%╙┼╩▐┴╙ ╬┴ ╦╥┴╙╬╧═ ╙═┼▌┼╬╔╔ $z$:
%
%$$
%\chi=\frac{c}{a(t_0) H_0}\int_0^z \frac{dz}{H(z)} =\frac{c}{a(t_0) H_0} \frac{1}{1-\alpha}
%[(1+z)^{1-\alpha}-1].
%$$ 
%
 where the observer is conveniently located at $\chi=0$.
For models with  $\alpha >1$, corresponding to decelerating universes, the integral
converges as $z \to\infty$.  It is well known that this indicates the existence of 
the particle horizon (particles with larger
$\chi $ can not be observed at the present moment).  For $\alpha <1$ the
particle horizon does not exist,
and the event horizon appears with the comoving coordinate:

$$
\chi_\mathrm{e.h.}=\frac{c}{a(t_0) H_0}\int_{-1}^0 \frac{dz}{H(z)}.
$$
This is the coordinate which light emitted now (from $\chi=0$)
will reach in the infinite future.  
Correspondingly, events which at present (according to cosmic time) occur on objects with 
$\chi > \chi_\mathrm{e.h.}$  will never be seen by the observer located at the
origin of the coordinate system.

%ф╠╤ ═╧─┼╠┼╩ ╙ $\alpha >1$, ╙╧╧╘╫┼╘╙╘╫╒└▌╔╚ ┌┴═┼─╠┼╬╬╧═╒ ╥┴╙█╔╥┼╬╔└, ╔╬╘┼╟╥┴╠
%╙╚╧─╔╘╙╤ ╨╥╔ $z \to\infty$, ▐╘╧, ╦┴╦ 
%╔┌╫┼╙╘╬╧, ╨╥╔╫╧─╔╘ ╦ ╬┴╠╔▐╔└ ╫ ═╧─┼╠╔ ╟╧╥╔┌╧╬╘┴ ▐┴╙╘╔├ (▐┴╙╘╔├┘ ╙ ┬\'╧╠╪█╔═
%$\chi $ ╫ ─┴╬╬┘╩ ═╧═┼╬╘ ╫╥┼═┼╬╔ ╬┴┬╠└─┼╬╔└
%╬┼ ─╧╙╘╒╨╬┘).  Ё╥╔ $\alpha <1$ ┼╟╧ ╬┼╘, ╬╧ ╨╧╤╫╠╤┼╘╙╤ ╟╧╥╔┌╧╬╘ ╙╧┬┘╘╔╩,
%╙╧╨╒╘╙╘╫╒└▌┴╤ ╦╧╧╥─╔╬┴╘┴ ╦╧╘╧╥╧╟╧ ─┴┼╘╙╤ ╞╧╥═╒╠╧╩:
%
%$$
%\chi_\mathrm{e.h.}=\frac{c}{a(t_0) H_0}\int_{-1}^0 \frac{dz}{H(z)},
%$$ 
%╘.┼.  ▄╘╧ ╦╧╧╥─╔╬┴╘┴, ╦╧╘╧╥╒└ ╔╙╨╒▌┼╬╬┘╩ ╙┼╩▐┴╙ ╠╒▐ ╙╫┼╘┴ ─╧╙╘╔╟╬┼╘ ┌┴
%┬┼╙╦╧╬┼▐╬╧┼ ╫╥┼═╤.  є╧╧╘╫┼╘╙╘╫┼╬╬╧, ╙╧┬┘╘╔╤, 
%╨╥╧╔╙█┼─█╔┼ ╬┴▐╔╬┴╤ ╙ ╘┼╦╒▌┼╟╧ ═╧═┼╬╘┴ ╨╧ ╦╧╙═╔▐┼╙╦╧═╒ ╫╥┼═┼╬╔ ╬┴ ╧┬▀┼╦╘┴╚ ╙
%$\chi > \chi_\mathrm{e.h.}$, ╬╔╦╧╟─┴ ╬┼ ┬╒─╒╘
%─╧╙╘╒╨╬┘ ╬┴┬╠└─┴╘┼╠└.  

It is instructive to compare how an object crosses the event horizon in an accelerating universe
with the same process in black holes (where one can define time at spatial infinity and
time on a free falling object, but there is no special time for the picture
``as a whole'', i.e. no analogue of the cosmic time).
Looking at the universe from ``god's point of view'' it is possible to
``see'' the event horizon at 
 $\chi = \chi_\mathrm{e.h.}$, and claim, for example, that:
 ``If the $\Lambda$CDM-model with the currently accepted values of cosmological parameters is
 true, then galaxies with
 $z>1.8$ are beyond the event horizon,'' -- the statement which is absolutely
 impossible in a description of free fall into a black hole.

A real cosmological observer has no access to the whole picture of 
the universe at the cosmic time ``now'', and the coordinate  
 $\chi = \chi_\mathrm{e.h.}$ has no special meaning for him/her.
 In this sense, the analogue of the infinite (from the position of a distant
 observer) process 
 $r \to r_\mathrm{g}$ of free fall into a black hole is the infinite process
 $z \to \infty$ seen by a cosmological observer.  Of course, the
 above-mentioned claim has a well-defined meaning without
any reference to the cosmic time ``now'' if we ``exchange'' the observer and the
source.  Namely, it means that any light emitted by us
now will never reach the galaxy
at $\chi = \chi_\mathrm{e.h.}$. (for more details, see \cite{Star, Varun}).  

%Ё╧╒▐╔╘┼╠╪╬╧ ╙╥┴╫╬╔╘╪ ╨┼╥┼╙┼▐┼╬╔┼ 
%╦╧╙═╧╠╧╟╔▐┼╙╦╧╟╧ ╟╧╥╔┌╧╬╘┴ ╙╧┬┘╘╔╩ ╧┬▀┼╦╘╧═ ╫ ╒╙╦╧╥┼╬╬╧ ╥┴╙█╔╥╤└▌┼╩╙╤
%ў╙┼╠┼╬╬╧╩ ╙ ─╥╒╟╔═ ``┴╥╚┼╘╔╨╔▐┼╙╦╔═'' ╨╥╧├┼╙╙╧═ ╫  
%яЇя --- ╨┴─┼╬╔┼═ ╧┬▀┼╦╘┴ ╫ █╫┴╥├█╔╠╪─╧╫╙╦╒└ ▐┼╥╬╒└ ─┘╥╒
%(╟─┼ ┼╙╘╪ ╫╥┼═╤ ┬┼╙╦╧╬┼▐╬╧
%╒─┴╠┼╬╬╧╟╧ ╬┴┬╠└─┴╘┼╠╤,  ╫╥┼═╤ ╙╫╧┬╧─╬╧ ╨┴─┴└▌┼╟╧ ╧┬▀┼╦╘┴, ╬╧ ╬┼╘ ╬╔╦┴╦╧╟╧
%╫┘─┼╠┼╬╬╧╟╧ ╫╥┼═┼╬╔ ─╠╤ ╦┴╥╘╔╬┘ ``╫ ├┼╠╧═'').
%ў┌╔╥┴╤ ╬┴ ╫╙┼╠┼╬╬╒└ ``╙ ╨╧┌╔├╔╔ ┬╧╟┴'', 
%═╧╓╬╧ ``╫╔─┼╘╪'' ╟╧╥╔┌╧╬╘ ╙╧┬┘╘╔╩ ╨╥╔ $\chi = \chi_\mathrm{e.h.}$
%╔ ─┼╠┴╘╪ ╒╘╫┼╥╓─┼╬╔╤ ╫╥╧─┼: ``х╙╠╔ $\Lambda$CDM-═╧─┼╠╪ ╙ ╙╧╫╥┼═┼╬╬┘═╔
%┌╬┴▐┼╬╔╤═╔ ╦╧╙═╧╠╧╟╔▐┼╙╦╔╚ ╨┴╥┴═┼╘╥╧╫ ╫┼╥╬┴, ╘╧ ╟┴╠┴╦╘╔╦╔ ╙ $z>1.8$
%╬┴╚╧─╤╘╙╤ ┌┴ ╟╧╥╔┌╧╬╘╧═ ╙╧┬┘╘╔╩'', --- ╦╧╘╧╥╧┼ ╙╧╫┼╥█┼╬╬╧
%╬┼═┘╙╠╔═╧ ╨╥╔ ╧╨╔╙┴╬╔╔ ╨┴─┼╬╔╤ ╘┼╠┴ ╫ ▐┼╥╬╒└ ─┘╥╒. ю┴┬╠└─┴╘┼╠╪, ╬┼ ╔═┼└▌╔╩ ─╧╙╘╒╨┴
%╦ ╨╧╠╬╧╩ ╦┴╥╘╔╬┼ ╫╙┼╠┼╬╬╧╩ ``╙┼╩▐┴╙'', ┼╙╘┼╙╘╫┼╬╬╧, 
%╬╔╦┴╦╧╟╧ ╟╧╥╔┌╧╬╘┴ ╬┼ ╫╔─╔╘, ╔ ─╠╤ ╬┼╟╧ ╙╧╨╒╘╙╘╫╒└▌┴╤
%╦╧╧╥─╔╬┴╘┴ $\chi = \chi_\mathrm{e.h.}$ ╬┼ ╫┘─┼╠┼╬┴ ╬╔▐┼═.
%  ў ▄╘╧═ ╙═┘╙╠┼ ┴╬┴╠╧╟╧═ ┬┼╙╦╧╬┼▐╬╧
%╥┴╙╘╤╬╒╘╧╟╧ (╘┼╧╥┼╘╔▐┼╙╦╔) ╨╥╧├┼╙╙┴ $r \to r_\mathrm{g}$ ─╠╤ ╨┴─┼╬╔╤ ╫ ▐┼╥╬╒└ ─┘╥╒
%╤╫╠╤┼╘╙╤ ╨╥╧├┼╙╙ $z \to \infty$. ы╧╬┼▐╬╧, ╒╨╧═╤╬╒╘╧╩ ╫┘█┼ ╞╥┴┌┼ ═╧╓╬╧ ╨╥╔─┴╘╪
%╙═┘╙╠
%╔ ┬┼┌ ╒╨╧═╔╬┴╬╔╤ ═╧═┼╬╘┴ ``╙┼╩▐┴╙'' ╨╧ ╦╧╙═╔▐┼╙╦╧═╒ ╫╥┼═┼╬╔, 
%┼╙╠╔ ``╨╧═┼╬╤╘╪ ═┼╙╘┴═╔'' ╔╙╘╧▐╬╔╦ ╔ ╬┴┬╠└─┴╘┼╠╤.
%с ╔═┼╬╬╧, ╧╬┴ ╧┌╬┴▐┴┼╘, ▐╘╧ ╔╙╨╒▌┼╬╬┘╩ ``╬┴═╔'' 
%╙┼╩▐┴╙ ╙╔╟╬┴╠ ╬╔╦╧╟─┴ ╬┼ ─╧╙╘╔╟╬┼╘ ▄╘╧╩ ╙┴═╧╩ ╟┴╠┴╦╘╔╦╔
%(╙═. ┬╧╠┼┼ ╨╧─╥╧┬╬╧┼ ╧┬╙╒╓─┼╬╔┼ ▄╘╧╟╧ ╫╧╨╥╧╙┴ ╫ \cite{Star}). 
%%Ї┴╦╓┼ ╬┼╦╧╘╧╥┘┼ ┴╙╨┼╦╘┘ ╙┴═╧╟╧ ╦╧╠╠┴╨╙┴ --- ╧┬╥┴┌╧╫┴╬╔╤
%%▐┼╥╬╧╩ ─┘╥┘ --- ╔═┼└╘ ┴╬┴╠╧╟╔╔ ╙ ╦╧╙═╧╠╧╟╔▐┼╙╦╔═╔ ╤╫╠┼╬╔╤═╔.

Now we address definitions of two different distance measures. One 
is valid in the ``god's perspective'', another is connected to the ``observer's
perspective''. 

%Ё┼╥┼╩─┼═ ╦ ╧╨╔╙┴╬╔└ ╥┴╙╙╘╧╤╬╔╩ ╫  ╘┼╚ ╙╠╒▐┴╤╚, 
%╨┼╥╫┘╩ ╔┌ ╦╧╘╧╥┘╚ ╧┬╥┴┌╬╧ ╬┴┌┘╫┴┼╘╙╤ ``╦┴╥╘╔╬╧╩ ┬╧╟┴'', 
%┴ ╫╘╧╥╧╩ --- ``╦┴╥╘╔╬╧╩ ╬┴┬╠└─┴╘┼╠╤''.

The proper distance is defined as  $d=a \chi$.
If we are interested in distances for $t=t_0$ (at the same moment of cosmic time
for all sources)
the scale factor in the equation should be equal to its present day value
$a(t_0)$. To obtain the proper distance at the moment of light emission we need the scale factor
at that time $a(t_\mathrm{em})$.

%є╧┬╙╘╫┼╬╬╧┼ ╥┴╙╙╘╧╤╬╔┼ ─╧ ╧┬▀┼╦╘┴ ╨╧ ╧╨╥┼─┼╠┼╬╔└ ╥┴╫╬╧:
% $d=a \chi$.
%ў ┌┴╫╔╙╔═╧╙╘╔ ╧╘ ╘╧╟╧, ╔╬╘┼╥┼╙╒┼═╙╤ ╠╔ ═┘ ╥┴╙╙╘╧╤╬╔┼═ ─╧ ╧┬▀┼╦╘┴ ``╙┼╩▐┴╙'' 
%(╧─╬╧╫╥┼═┼╬╬╧ ╙ ╬┴═╔ ╨╧ ╦╧╙═╔▐┼╙╦╧═╒ ╫╥┼═┼╬╔)
%╔╠╔ ╘╧╟─┴, ╦╧╟─┴ ╧┬▀┼╦╘ ╔╙╨╒╙╘╔╠ ╨╥╔╬╔═┴┼═┘╩ ╬┴═╔ ╙┼╩▐┴╙ ╙╫┼╘, 
%╨╧─ ═┴╙█╘┴┬╬┘═ ╞┴╦╘╧╥╧═ ╙╠┼─╒┼╘ ╨╧╬╔═┴╘╪
%┼╟╧ ╬┘╬┼█╬┼┼ ┌╬┴▐┼╬╔┼ $a(t_0)$ 
%╔╠╔ ┌╬┴▐┼╬╔┼ ╫ ═╧═┼╬╘ ╔╙╨╒╙╦┴╬╔╤ ╙╫┼╘┴,
%$a(t_\mathrm{em})$.

The general formula for the proper distance at the present moment, $t_0$, is the following:
$$
d=\frac{c}{(1-\alpha) H_0} [(1+z)^{1-\alpha}-1].
$$
%(╞╧╥═┴╠╪╬╧ ╙╧╫╨┴─┴└▌┼┼ ╙ ╫╫╧─╔═┘═ ╫ ╬┴┬╠└─┴╘┼╠╪╬╧╩ ╦╧╙═╧╠╧╟╔╔ ╥┴╙╙╘╧╤╬╔┼═ ╨╧ ╙╧┬╙╘╫┼╬╬╧═╒ ─╫╔╓┼╬╔└) 
This distance is useful in order to imagine the general structure of the universe we live in.
It grows monotonically with increasing redshift, tending to a finite
value if  $z \to \infty$ for
 $\alpha > 1$.  
This gives us an intuitively clear picture of a finite distance to the particle horizon.
 
%є╧┬╙╘╫┼╬╬╧┼ ╥┴╙╙╘╧╤╬╔┼ ─╧ ╧┬▀┼╦╘┴ ``╙┼╩▐┴╙'' ═╧╓╬╧ ┌┴╨╔╙┴╘╪ ╦┴╦:
%
%$$
%d=\frac{c}{(1-\alpha) H_0} [(1+z)^{1-\alpha}-1].
%$$
%%(╞╧╥═┴╠╪╬╧ ╙╧╫╨┴─┴└▌┼┼ ╙ ╫╫╧─╔═┘═ ╫ ╬┴┬╠└─┴╘┼╠╪╬╧╩ ╦╧╙═╧╠╧╟╔╔ ╥┴╙╙╘╧╤╬╔┼═ ╨╧ ╙╧┬╙╘╫┼╬╬╧═╒ ─╫╔╓┼╬╔└) 
%я╨╥┼─┼╠┼╬╬╧┼ ╘┴╦╔═ ╧┬╥┴┌╧═ ╥┴╙╙╘╧╤╬╔┼
%╨╧╠┼┌╬╧ ─╠╤ ╙╧┌─┴╬╔╤ ╫ ╟╧╠╧╫┼ ╬┴┬╠└─┴╘┼╠╤ ┬╧╠┼┼ ╔╠╔ ═┼╬┼┼ ╨╧╬╤╘╬╧╟╧ ╧┬╥┴┌┴
%╫╙┼╠┼╬╬╧╩, ╫ ╦╧╘╧╥╧╩ ╧╬ ╓╔╫┼╘. 
%я╬╧ ═╧╬╧╘╧╬╬╧ ╥┴╙╘┼╘ ╙ ╥╧╙╘╧═ ╦╥┴╙╬╧╟╧ ╙═┼▌┼╬╔╤, 
%╙╘╥┼═╤╙╪ ╫ ╙╠╒▐┴┼ $\alpha > 1$ ╨╥╔ $z \to \infty$ ╦ ╦╧╬┼▐╬╧═╒ ╨╥┼─┼╠╒,
% ▐╘╧ ╙╧╟╠┴╙╒┼╘╙╤ ╙ ╔╬╘╒╔╘╔╫╬╧╩ ╦┴╥╘╔╬╧╩ ╦╧╬┼▐╬╧╟╧ ╥┴╙╙╘╧╤╬╔╤ ─╧
%╟╧╥╔┌╧╬╘┴  ▐┴╙╘╔├.

In what follows, we will mostly consider another possiblity, 
which is related to the picture which an observer can really see.
The proper distance to an object at the moment of light emission,
 $d(t_\mathrm{em})$, 
coincides with the well-known (and sometimes directly observable) angular
distance, $d_{\Theta}$, defined as $d_{\Theta}=D/\delta$, where $D$ is the diameter of 
an emitter, and $\delta$ is its observed angular diameter.  
%{\bf Lesha: $d_{\Theta}$ needs to be defined here.}

For a perfect fluid with parameter $\alpha$ we have:

$$
%\begin{equation} 
d_{\Theta}=\frac{c}{H_0}\frac{1}{1-\alpha}[(1+z)^
{1-\alpha}-1]\frac{1}{(1+z)}.
$$
%\end{equation}

%я─╬┴╦╧ ╬┴═ ╙┼╩▐┴╙ ┬╧╠┼┼ ╔╬╘┼╥┼╙┼╬ ─╥╒╟╧╩ ╫┴╥╔┴╬╘ ╧╨╥┼─┼╠┼╬╔╤ ╥┴╙╙╘╧╤╬╔╤, 
%╘.╦. ╧╬ ╧╘╥┴╓┴┼╘  ╘╧, ▐╘╧ ╫╧╙╨╥╔╬╔═┴┼╘ ╬┴┬╠└─┴╘┼╠╪. 
%%╙╫╧╩╙╘╫╧ ╙╧┬╙╘╫┼╬╬╧ ╬┴┬╠└─┴┼═╧╟╧ ╧┬▀┼╦╘┴. 
%є╧┬╙╘╫┼╬╬╧┼
%╥┴╙╙╘╧╤╬╔┼ ─╧ ╔╙╘╧▐╬╔╦┴ ╫ ═╧═┼╬╘ ╔┌╠╒▐┼╬╔╤ $d(t_\mathrm{em})$ ╙╧╫╨┴─┴┼╘ ╙ ╚╧╥╧█╧ ╔┌╫┼╙╘╬┘═ 
%╫ ┴╙╘╥╧╞╔┌╔╦┼ (╔ ╫ ╥╤─┼ ╫┴╓╬┘╚ ╙╠╒▐┴┼╫ ╬┼╨╧╙╥┼─╙╘╫┼╬╬╧ ╬┴┬╠└─┴┼═┘═) ╥┴╙╙╘╧╤╬╔┼═ ╨╧ ╒╟╠╧╫╧═╒ ─╔┴═┼╘╥╒ 
%$d_{\Theta}$, ╧╨╔╙┘╫┴└▌┼═╒ ╔┌═┼╬┼╬╔┼ ╒╟╠╧╫╧╟╧
%╥┴┌═┼╥┴ ╧┬▀┼╦╘┴ ╙ ╞╔╦╙╔╥╧╫┴╬╬┘═ ╠╔╬┼╩╬┘═ ╥┴┌═┼╥╧═ ╙ ╥╧╙╘╧═ ╦╥┴╙╬╧╟╧ ╙═┼▌┼╬╔╤. ф╠╤ 
%┬┴╥╧╘╥╧╨╬╧╩ ═┴╘┼╥╔╔ ╫ ╧┬▌┼═ ╙╠╒▐┴┼:
%$$
%%\begin{equation} 
%d_{\Theta}=\frac{c}{H_0}\frac{1}{1-\alpha}[(1+z)^
%{1-\alpha}-1]\frac{1}{(1+z)}.
%$$
%%\end{equation}

In contrast to the present day proper distance (i.e., distance ``now''), 
the dependence of the angular distance 
on the redshift is not monotonic for most 
realistic cosmological models (if we do not consider some exotics, only in de Sitter model with
 $\alpha=0$ the function $d_{\Theta}(z)$
has no local maximum). For example, for a dust-dominated universe 
($p=0$, so $\alpha=3/2$) $d_{\Theta}$ reaches its maximum at 
 $z=5/4$ which is well within the currently observable universe.
That is why it may be hard to believe that this quantity is the most 
relevant characteristic of the distance.
However, from the point of view of observations it is exactly what we need. 
Remember, for example, the  observational proof
that our universe is close to the flat one, which is
based on calculations of the angular distance to the surface of last
scattering.
It is true (though it may sound strange) that spots on the last scattering surface  
 ($z \sim 1100$) seen in the CMB temperature maps were situated (at the time of
emission) at the same distance 
(about 13 Mpc) as 
some near-by galaxy with $z \approx 0.003$
(of course, the moment of emission for this galaxy was much later with respect to
cosmic time). 
 \footnote{It is convenient to use on-line cosmological calculators for such estimates,
for example, the one by Ned Wright
http://www.astro.ucla.edu/$\sim$wright/CosmoCalc.html.}

%ў ╧╘╠╔▐╔┼ ╧╘ ╥┴╙╙╘╧╤╬╔╤ ╨╧ ╙╧┬╙╘╫┼╬╬╧═╒ ─╫╔╓┼╬╔└ (╙═. ╬╔╓┼), 
%┌┴╫╔╙╔═╧╙╘╪ ╒╟╠╧╫╧╟╧ ╥┴╙╙╘╧╤╬╔╤ ╧╘ ╦╥┴╙╬╧╟╧
%╙═┼▌┼╬╔╤ ╬┼═╧╬╧╘╧╬╬┴ ─╠╤ ┬╧╠╪█╔╬╙╘╫┴ ╥┼┴╠╔╙╘╔▐┼╙╦╔╚ ╦╧╙═╧╠╧╟╔▐┼╙╦╔╚ ═╧─┼╠┼╩
%(╙╧┬╙╘╫┼╬╬╧, ┼╙╠╔ ╬┼ ╨╥╔╫╠┼╦┴╘╪ ▄╦┌╧╘╔╦╒, ╠╔█╪ ╫ ═╔╥┼ ─┼ є╔╘╘┼╥┴, ─╠╤
%╦╧╘╧╥╧╟╧ $\alpha=0$, ╒ ╞╒╬╦├╔╔ $d_{\Theta}(z)$
%╬┼╘ ╠╧╦┴╠╪╬╧╟╧ ═┴╦╙╔═╒═┴). ю┴╨╥╔═┼╥, ─╠╤ ▐╔╙╘╧ ╨┘╠┼╫╧╩ ў╙┼╠┼╬╬╧╩ 
%(╫ ╬┼╩ $p=0$, ╘.┼. $\alpha=3/2$)
%╧╬┴ ─╧╙╘╔╟┴┼╘ ═┴╦╙╔═╒═┴
%╨╥╔ ╬┼ ╘┴╦╧═ ╒╓ ╔ ┬╧╠╪█╧═ ┌╬┴▐┼╬╔╔ $z=5/4$.
%Ё╧▄╘╧═╒ ╬┴ ╨┼╥╫┘╩ ╫┌╟╠╤─, ┌┴ ▄╘╧╩ ╞╒╬╦├╔┼╩ ╘╤╓┼╠╧ ╨╥╔┌╬┴╘╪
%╬┴╔┬╧╠┼┼ ╠╧╟╔▐╬╒└ ═┼╥╒ ╥┴╙╙╘╧╤╬╔╤. я─╬┴╦╧ ╙ ╘╧▐╦╔ ┌╥┼╬╔╤ ╬┴┬╠└─┼╬╔╩ ▄╘╧ ╥┴╙╙╘╧╤╬╔┼ ╨╥┼─╙╘┴╫╠╤┼╘
%┬╧╠╪█╧╩ ╔╬╘┼╥┼╙.  ў ▐┴╙╘╬╧╙╘╔, 
%─╧╦┴┌┴╘┼╠╪╙╘╫╧ ╘╧╟╧, ▐╘╧ ╬┴█┴ ╫╙┼╠┼╬╬┴╤ ┬╠╔┌╦┴ ╦ ╨╠╧╙╦╧╩, ╨╥╔ ╨╧═╧▌╔  ╥┼╠╔╦╘╧╫╧╟╧ ╔┌╠╒▐┼╬╔╤ 
%╧╙╬╧╫┴╬╧ ╔═┼╬╬╧ ╬┴ ╧╨╥┼─┼╠┼╬╔╔ ╒╟╠╧╫╧╟╧ ╥┴╙╙╘╧╤╬╔╤
%─╧ ╨╧╫┼╥╚╬╧╙╘╔ ╨╧╙╠┼─╬┼╟╧ ╥┴╙╙┼╤╬╔╤. щ, ╦┴╦ ▄╘╧ ╬╔ ╨╧╦┴╓┼╘╙╤ ╬┴ ╨┼╥╫┘╩ ╫┌╟╠╤─
%╙╘╥┴╬╬┘═, ╨╤╘╬┴ ╬┴ ╨╧╫┼╥╚╬╧╙╘╔ ╨╧╙╠┼─╬┼╟╧ ╥┴╙╙┼╤╬╔╤ ($z \sim 1100$), ╦╧╘╧╥┘┼
%═┘ ╫╔─╔═ ╬┴ ╦┴╥╘┴╚ ╥┴╙╨╥┼─┼╠┼╬╔╤ ╘┼═╨┼╥┴╘╒╥┘ ╥┼╠╔╦╘╧╫╧╟╧ ╔┌╠╒▐┼╬╔╤, ╬┴╚╧─╔╠╔╙╪ ╫
%═╧═┼╬╘ ╔┌╠╒▐┼╬╔╤ ╨╥╔╬╔═┴┼═╧╟╧ ╙┼╩▐┴╙ ╙╫┼╘╧╫╧╟╧ ╙╔╟╬┴╠┴ ╬┴ ╘╧═ ╓┼ ╥┴╙╙╘╧╤╬╔╔
%(╧╦╧╠╧ 13 э╨╦, ╘╧▐╬╧┼ ┌╬┴▐┼╬╔┼ ┌┴╫╔╙╔╘ ╧╘ ╨┴╥┴═┼╘╥╧╫ ═╧─┼╠╔) ▐╘╧ ╔ ╦┴╦┴╤-╬╔┬╒─╪ ┬╠╔┌╦┴╤ 
%╟┴╠┴╦╘╔╦┴ ╙ $z \approx 0.003$ 
%(╨╧╬╤╘╬╧, ▐╘╧  ▄╘╧╘ ═╧═┼╬╘ ╔┌╠╒▐┼╬╔╤ ╨╥╔╬╔═┴┼╟╧╟╧ ╙┼╩▐┴╙ ╙╫┼╘┴ ╨╧ ╦╧╙═╔▐┼╙╦╧═╒ ╫╥┼═┼╬╔
%─╠╤ ╟┴╠┴╦╘╔╦╔ ╬┴╙╘╒╨╔╠ ┌╬┴▐╔╘┼╠╪╬╧ ╨╧┌╓┼). \footnote{э╬╧╟╔┼ ▐╔╙╠┼╬╬┘┼ ╧├┼╬╦╔
%╒─╧┬╬╧ ─┼╠┴╘╪ ╙ ╨╧═╧▌╪└ ╧╬-╠┴╩╬╧╫┘╚ ╦╧╙═╧╠╧╟╔▐┼╙╦╔╚ ╦┴╠╪╦╒╠╤╘╧╥╧╫, ╬┴╨╥╔═┼╥,
%╔╙╨╧╠╪┌╒╤
%╙╧┌─┴╬╬┘╩ ю▄─╧═ Є┴╩╘╧═
%http://www.astro.ucla.edu/$\sim$wright/CosmoCalc.html.}

As light trajectories in a flat universe are straight lines (see Fig. 1), the fact that
angular size of distant objects grows with growing redshift
has just this reason --- objects with larger redshifts were at the time of
emission closer to us than objects with lower
redshifts (for $z$ large enough).  Of course, if we use $z$ itself to describe
how distant an object is from us 
(as it often happens), such 
situation does not occur.  However, if our goal is to introduce a
meaningful measure of distance in the picture
directly seen by an observer, it is the angular distance that plays this role.  

%Ё╧╙╦╧╠╪╦╒ ╘╥┴┼╦╘╧╥╔╔ ╙╫┼╘╧╫┘╚ ╠╒▐┼╩
%╫ ╨╠╧╙╦╧═ ═╔╥┼ ╧╙╘┴└╘╙╤ ╨╥╤═┘═╔, ╒╫┼╠╔▐┼╬╔┼ ╒╟╠╧╫╧╟╧ ╥┴┌═┼╥┴ ╧┬▀┼╦╘┴ ╙ ╥╧╙╘╧═ 
%╦╥┴╙╬╧╟╧ ╙═┼▌┼╬╔╤ ╧╘╥┴╓┴┼╘ ╔═┼╬╬╧ ▄╘╧╘ ╞┴╦╘ --- ╧┬▀┼╦╘┘ ╙ ┬╧╠╪█╔═ ╦╥┴╙╬┘═ ╙═┼▌┼╬╔┼═
%┬┘╠╔ ╫ ═╧═┼╬╘ ╔┌╠╒▐┼╬╔╤ ┬╠╔╓┼ ╦ ╬┴═, ▐┼═ ╧┬▀┼╦╘┘ ╙ ═┼╬╪█╔═ ╦╥┴╙╬┘═ ╙═┼▌┼╬╔┼═
%(─╠╤ ─╧╙╘┴╘╧▐╬╧ ┬╧╠╪█╔╚ $z$), ╙═. ╥╔╙.\ref{sketch2}. 
%х╙╘┼╙╘╫┼╬╬╧, ╨┼╥┼╚╧─ ╦ $z$ ╦┴╦ ``═┼╥┼ ╒─┴╠┼╬╬╧╙╘╔''
%╔╙╘╧▐╬╔╦╧╫ ╙╫┼╘┴ ╘╒╘ ╓┼ ╫╧╙╙╘┴╬┴╫╠╔╫┴┼╘ ╨╥╔╫┘▐╬╒└ ╦┴╥╘╔╬╒. я─╬┴╦╧, ┼╙╠╔ ╥┼▐╪
%╔─┼╘ ╧ ╨╥╔─┴╬╔╔ ╙═┘╙╠┴ ╥┴╙╙╘╧╤╬╔╤═ ─╧ ─┴╠┼╦╔╚ ╧┬▀┼╦╘╧╫
%╫ ╦┴╥╘╔╬┼ ╫╙┼╠┼╬╬╧╩, ╬┼╨╧╙╥┼─╙╘╫┼╬╬╧ ╞╔╦╙╔╥╒┼═╧╩ ╬┴┬╠└─┴╘┼╠┼═,
%╔═┼╬╬╧ ╒╟╠╧╫╧┼ ╥┴╙╙╘╧╤╬╔┼ ╬┴╔╠╒▐█╔═ ╧┬╥┴┌╧═ ╫┘╨╧╠╬╤┼╘ ▄╘╒ ╥╧╠╪.  
 
In the final part of this section we briefly remind 
definitions of several other cosmological distances.

%ў ┌┴╦╠└▐┼╬╔┼ ╥┴┌─┼╠┴ ─╠╤ ╨╧╠╬╧╘┘ ╦┴╥╘╔╬┘ ╦╥┴╘╦╧ ╧╙╘┴╬╧╫╔═╙╤ ╬┴ ╞╧╘╧═┼╘╥╔▐┼╙╦╧═ ╥┴╙╙╘╧╤╬╔╔,
%$d_\mathrm{ph}$ ╔ ╥┴╙╙╘╧╤╬╔╔ ╨╧ ╙╧┬╙╘╫┼╬╬╧═╒ ─╫╔╓┼╬╔└, $d_\mathrm{pm}$.

The photometric distance is rather popular in observational cosmology. It is defined
as:
%\begin{equation}
$$
d_\mathrm{ph}=(L/4\pi f)^{1/2}=a^2(t_0)\frac{\chi}{a(t_\mathrm{em})},
$$
%\end{equation}
where $L$ is the luminosity of an object, $f$ is the observed flux. 
 Note, that the photometric distance diverges at the event horizon (if it exists in the
model). However, this is because of the energy dilution due to redshift 
encoded in the definition of this parameter.  

%ц╧╘╧═┼╘╥╔▐┼╙╦╧┼ ╥┴╙╙╘╧╤╬╔┼ ╫┼╙╪═┴ ╨╧╨╒╠╤╥╬╧ ╫ ╬┴┬╠└─┴╘┼╠╪╬╧╩ ╦╧╙═╧╠╧╟╔╔.
%я╬╧ ╧╨╥┼─┼╠╤┼╘╙╤ ╦┴╦:
%%\begin{equation}
%$$
%d_\mathrm{ph}=(L/4\pi f)^{1/2}=a^2(t_0)\frac{\chi}{a(t_\mathrm{em})},
%$$
%%\end{equation}
%╟─┼ $L$ --- ╙╫┼╘╔═╧╙╘╪ ╔╙╘╧▐╬╔╦┴, ┴ $f$ --- ╨╧╘╧╦ ╨╥╔╬╔═┴┼═╧╟╧ ╔┌╠╒▐┼╬╔╤. 
%я╘═┼╘╔═, ▐╘╧
%╞╧╘╧═┼╘╥╔▐┼╙╦╧┼ ╥┴╙╙╘╧╤╬╔┼ ╧┬╥┴▌┴┼╘╙╤ ╬┴ ╟╧╥╔┌╧╬╘┼ ╙╧┬┘╘╔╩ 
%(┼╙╠╔ ╘┴╦╧╩ ╨╥╔╙╒╘╙╘╫╒┼╘) ╫ ┬┼╙╦╧╬┼▐╬╧╙╘╪, ╬╧ ▄╘╧ ╙╫╤┌┴╬╧ ╙ ┌┴╦╧─╔╥╧╫┴╬╬┘═ ╫ ┼╟╧
%╧╨╥┼─┼╠┼╬╔╔ ╨┴─┼╬╔┼═ ╔╬╘┼╬╙╔╫╬╧╙╘╔
%╔┌╠╒▐┼╬╔╤ ╔┌-┌┴ ╦╥┴╙╬╧╟╧ ╙═┼▌┼╬╔╤, ┴ ╬┼ ╙ ╥┴╙╙╘╧╤╬╔┼═ ╦┴╦ ╘┴╦╧╫┘═. 

The proper motion distance is rather interesting because it formally coinsides with the proper
distance for the present moment: $d_\mathrm{pm}=a(t_0) \chi$. 
Currently there are no effective methods to determine this distance.
In principle, they can appear in connection to studies of jets in distant
sources.

%Є┴╙╙╘╧╤╬╔┼ ╨╧ ╙╧┬╙╘╫┼╬╬╧═╒ ─╫╔╓┼╬╔└ ╨╥╔═┼▐┴╘┼╠╪╬╧ ╘┼═, ▐╘╧ ╙╧╫╨┴─┴┼╘ ╙
%╙╧┬╙╘╫┼╬╬┘═ ╥┴╙╙╘╧╤╬╔┼═ ╫ ═╧═┼╬╘ ╬┴┬╠└─┼╬╔╤: $d_\mathrm{pm}=a(t_0) \chi$. 
%ю┴ ╬┴╙╘╧╤▌╔╩ ═╧═┼╬╘ ╙ ╘╧▐╦╔ ┌╥┼╬╔╤ ╬┴┬╠└─┼╬╔╩ ╬┼╘ ╚╧╥╧█╔╚ ═┼╘╧─╔╦
%╧╨╥┼─┼╠┼╬╔╤ ▄╘╧╟╧ ╨┴╥┴═┼╘╥┴. Ё╧╘┼╬├╔┴╠╪╬╧ ╧╬╔ ═╧╟╒╘ ╨╧╤╫╔╘╪╙╤ ╫ ╙╫╤┌╔ ╙
%╔┌╒▐┼╬╔┼═ ─╓┼╘╧╫ ─┴╠┼╦╔╚ ╔╙╘╧▐╬╔╦╧╫. 

It should be noted that the popularity of photometric distance in contemporary
observational cosmology
is related to the existence of ``standard candles'', but not to some
special role of this distance in theoretical
models.  The angular distance is currently not so popular, however, the
situation may change with appearence
of a ``standard ruler''.  Some time ago it was proposed to use a
characteristic distance
scale of barionic acoustic oscillation for this purpose
\cite{bh2010}.
This proposal have been already used in a several studies 
 \cite{percival, blake, busca}. These authors used  data on
hundreds of thousands of galaxies and quasars 
to estimate basic cosmological parameters.

%є╠┼─╒┼╘ ╧╘═┼╘╔╘╪, ▐╘╧ ╫ ╬┴╙╘╧╤▌╔╩ ═╧═┼╬╘ ╙ ╘╧▐╦╔ ┌╥┼╬╔╤ ╬┴┬╠└─┴╘┼╠╪╬╧╩
%╦╧╙═╧╠╧╟╔╔ ╞╧╘╧═┼╘╥╔▐┼╙╦╧┼ ╥┴╙╙╘╧╤╬╔┼ ╤╫╠╤┼╘╙╤ ┬╧╠┼┼ ▐┴╙╘╧ ╔╙╨╧╠╪┌╒┼═┘═
%┬╠┴╟╧─┴╥╤ ╬┴╠╔▐╔└ ``╙╘┴╬─┴╘╬┘╚ ╙╫┼▐'' (╘.┼., ┴╙╘╥╧╬╧═╔▐┼╙╦╔╚ ╔╙╘╧▐╬╔╦╧╫ ╙
%╔┌╫┼╙╘╬╧╩ ╙╫┼╘╔═╧╙╘╪└), ┴ ╬┼ ╨╧ ╨╥╔▐╔╬┼ ╫┘─┼╠┼╬╬╧╙╘╔ ▄╘╧╟╧
%╥┴╙╙╘╧╤╬╔╤ ╫ ╘┼╧╥┼╘╔▐┼╙╦╔╚ ═╧─┼╠╤╚. ї╟╠╧╫╧┼ ╥┴╙╙╘╧╤╬╔┼ ╨╧╦┴ ╔╟╥┴┼╘ ═┼╬╪█╒└ ╥╧╠╪, ╬╧ ╙╔╘╒┴├╔╤ ═╧╓┼╘ ╔┌═┼╬╔╘╪╙╤ ╨╧╙╠┼
%╨╧╤╫╠┼╬╔╤ ``╙╘┴╬─┴╥╘╬╧╩ ╠╔╬┼╩╦╔''. 
%ю┼╦╧╘╧╥╧┼ ╫╥┼═╤ ╬┴┌┴─  ╨╧╤╫╔╠╔╙╪ ╨╥┼─╠╧╓┼╬╔╤ ╔╙╨╧╠╪┌╧╫┴╘╪ ─╠╤ ▄╘╧╟╧
%%ї╓┼ ╨╧╤╫╔╠╔╙╪ ╨╥┼─╠╧╓┼╬╔╤ ╔╙╨╧╠╪┌╧╫┴╘╪ ─╠╤ ▄╘╧╟╧
%╚┴╥┴╦╘┼╥╬┘╩ ═┴╙█╘┴┬ ┬┴╥╔╧╬╬┘╚ ┴╦╒╙╘╔▐┼╙╦╔╚ ╧╙├╔╠╠╤├╔╩ \cite{bh2010}.
%ў ╥╤─┼ ╥┴┬╧╘ ╘┴╦╧╩ ╨╧─╚╧─ ╒╓┼ ╔╙╨╧╠╪┌╧╫┴╬ \cite{percival, blake, busca}.
%щ╙╨╧╠╪┌╒╤ ─┴╬╬┘┼ ╨╧ ╥┴╙╨╥┼─┼╠┼╬╔└ ┬╧╠╪█╧╟╧ ▐╔╙╠┴ --- ─┼╙╤╘╦╔ ╔ ╙╧╘╬╔ ╘┘╙╤▐,
%---
%╦╫┴┌┴╥╧╫ ╔╠╔ ╟┴╠┴╦╘╔╦, ╙╧╧╘╫┼╘╙╘╫┼╬╬╧,
%┴╫╘╧╥┘ ╨╧╠╒▐╔╠╔ ╧├┼╬╦╔ ╧╙╬╧╫╬┘╚ ╦╧╙═╧╠╧╟╔▐┼╙╦╔╚ ╨┴╥┴═┼╘╥╧╫.

We would like to underline once again that a special role of the angular
distance,
in our opinion, is not 
related to the quality of currently available astrophysical data, but follows
from the fact that this distance
coincides with the fundamental theoretical quantity --- proper distance at
the moment of emission.  

%э┘ ╚╧╘╔═ ╨╧─▐┼╥╦╬╒╘╪, ▐╘╧   ╫┘─┼╠┼╬╬╧╙╘╪  ╒╟╠╧╫╧╟╧
%╥┴╙╙╘╧╤╬╔╤ ╫ ╬┴█┼═ ╥┴╙╙═╧╘╥┼╬╔╔ ╬┼ ╙╫╤┌┴╬┴ ╙ ╘┼╦╒▌┼═ ╒╥╧╫╬┼═ 
%┴╙╘╥╧╞╔┌╔▐┼╙╦╔╚ ┌╬┴╬╔╩, ┴ ╨╥╧╔╙╘┼╦┴┼╘ ╔┌ ╘╧╟╧, ▐╘╧ ╧╬╧ ╙╧╫╨┴─┴┼╘ ╙ ╞╒╬─┴═┼╬╘┴╠╪╬┘═ ╧┬▀┼╦╘╧═
%╘┼╧╥╔╔ --- ╙╧┬╙╘╫┼╬╬┘═ ╥┴╙╙╘╧╤╬╔┼═ ╫ ═╧═┼╬╘
%╔┌╠╒▐┼╬╔╤ ╨╥╔╬╔═┴┼═╧╟╧ ╙┼╩▐┴╙ ╙╔╟╬┴╠┴. 

Different distances are related to each other as follows:
%\begin{equation}
$$
d_\Theta=a(t_\mathrm{em})\chi=\frac{d_\mathrm{pm}}{(1+z)}=\frac{d_\mathrm{ph}}{(1+z)^2}.
$$

%Є┴┌╬┘┼ ╥┴╙╙╘╧╤╬╔╤ ╙╫╤┌┴╬┘ ─╥╒╟ ╙ ─╥╒╟╧═ ╙╠┼─╒└▌╔═ ╧┬╥┴┌╧═:
%%\begin{equation}
%$$
%d_\Theta=a(t_\mathrm{em})\chi=\frac{d_\mathrm{pm}}{(1+z)}=\frac{d_\mathrm{ph}}{(1+z)^2}.
%$$
%%\end{equation}
%
%%\newpage 

\begin{figure}
\resizebox{\hsize}{!}
{\includegraphics[angle=90]{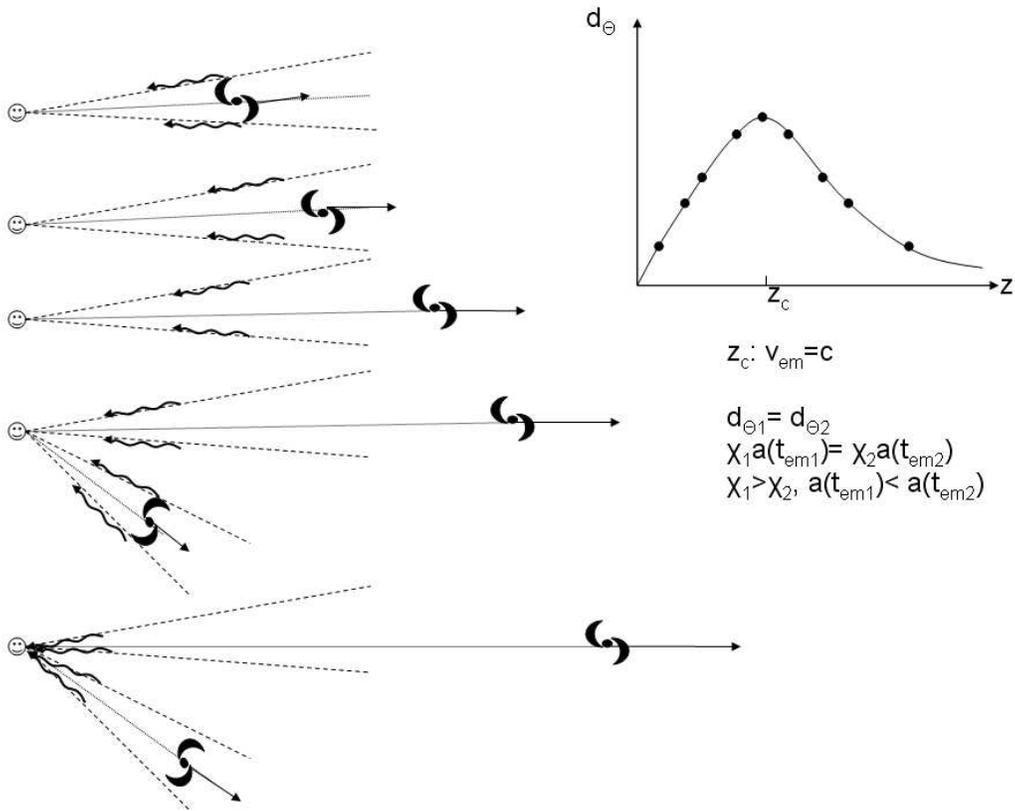}}
\caption{This sketch illustrates that objects with the same angular
distance, but different comoving coordinates, form pairs relative to the
maximum of the function $d_\Theta(z)$. This maximum corresponds to the
redshift at which the recession velocity at the moment of emission is equal
to the velocity of light. Light of the more distant galaxy in a pair at first recedes,  
but then starts to approach the observer, still the angle between rays from
opposite sides of the galaxy towards the observer remains constant. 
%%є╚┼═┴╘╔▐╬╧ ╨╧╦┴┌┴╬╧, ▐╘╧ ╧┬▀┼╦╘┘ ╙ ╧─╔╬┴╦╧╫┘═╔ ╒╟╠╧╫┘═╔
%╥┴╙╙╘╧╤╬╔╤═╔, ╬╧ ╥┴┌╬┘═╔ ╙╧╨╒╘╙╘╫╒└▌╔═╔ ╦╧╧╥─╔╬┴╘┴═╔, ╧┬╥┴┌╒└╘ ╨┴╥┘
%╧╘╬╧╙╔╘┼╠╪╬╧ ═┴╦╙╔═╒═┴ ╫ ┌┴╫╔╙╔═╧╙╘╔ $d_\Theta$ ╧╘ $z$. 
%э┴╦╙╔═╒═ ╙╧╧╘╫┼╘╙╘╫╒┼╘ ╦╥┴╙╬╧═╒ ╙═┼▌┼╬╔└, ╬┴ ╦╧╘╧╥╧═ ╙╦╧╥╧╙╘╪ ╬┴ ═╧═┼╬╘
%╔┌╠╒▐┼╬╔╤ ╥┴╫╬┴ ╙╦╧╥╧╙╘╔ ╙╫┼╘┴. є╫┼╘╧╫┘┼ ╠╒▐╔ ╧╘ ┬╧╠┼┼ ─┴╠┼╦╧╩ ╟┴╠┴╦╘╔╦╔
%╔┌╬┴▐┴╠╪╬╧ ╒─┴╠╤└╘╙╤ ╧╘ ╬┴┬╠└─┴╘┼╠╤, ╔ ╠╔█╪ ╨╧╘╧═ ╬┴▐╔╬┴└╘ ╨╥╔┬╠╔╓┴╘╪╙╤, ╬╧
%╒╟╧╠ ═┼╓─╒ ╬╔═╔ ╙╧╚╥┴╬╤┼╘╙╤.
} 
\label{sketch2}
\end{figure}

%\newpage

\section{Velocities in cosmology}

Before we address the problem of definition of the velocity characterizing the Hubble flow, 
it is necessary to note that the existence of different
velocities with different meanings is rather typical in the General relativity. 
A classical example of this situation is the description of
free fall into a black hole.  From the point of  view of an observer at spatial
infinity the motion of falling objects
is initially accelerated, then, after reaching its maximum the coordinate velocity
(which in the case of free fall from
infinity with zero initial velocity is equal to $2c/[3/\sqrt{3}$]) is
decreasing \cite{okun}.  It is clear that this
coordinate velocity defined as the ratio of distance to the time
interval measured at the infinity
is important only for description of an observed picture of free fall.  It
is useless when we describe processes
in the vicinity of the black hole itself.  Our goal is to find an analogue
for such velocity in the case of the Hubble flow.  

%я┬╥┴▌┴╤╙╪ ╦ ╫╧╨╥╧╙╒ ╧ ╙╦╧╥╧╙╘╔ ╚┴┬┬╠╧╫╙╦╧╟╧ ╨╧╘╧╦┴, ╬┼╧┬╚╧─╔═╧ ┌┴═┼╘╔╘╪,
%▐╘╧ ╬┴╠╔▐╔┼ ╥┴┌╬┘╚ ╫╔─╧╫ ╙╦╧╥╧╙╘┼╩, ╨╥╔═┼╬╔═┘╚ ╫ ╥┴┌╠╔▐╬┘╚ ╙╔╘╒┴├╔╤╚, ╫╧╧┬▌┼
%╘╔╨╔▐╬╧ ─╠╤ яЇя. ы╠┴╙╙╔▐┼╙╦╔═ ╨╥╔═┼╥╧═ ╤╫╠╤┼╘╙╤ ╧╨╔╙┴╬╔┼ ╙╫╧┬╧─╬╧╟╧ ╨┴─┼╬╔╤
%╧┬▀┼╦╘┴ ╫ █╫┴╥├█╔╠╪─╧╫╙╦╒└ ▐┼╥╬╒└ ─┘╥╒, ╦╧╘╧╥╧┼ ╙ ╘╧▐╦╔ ┌╥┼╬╔╤ ╬┴┬╠└─┴╘┼╠╤
%╬┴ ┬┼╙╦╧╬┼▐╬╧╙╘╔ ╫╔─╔╘╙╤ ╫ ╬┴▐┴╠┼ ╒╙╦╧╥┼╬╬┘═, ┴ ╨╧╘╧═, ╔┌-┌┴ ╟╥┴╫╔╘┴├╔╧╬╬╧╟╧ 
%┌┴═┼─╠┼╬╔╤ ╫╥┼═┼╬╔, ┌┴═┼─╠┼╬╬┘═ ╨╧╙╠┼ ─╧╙╘╔╓┼╬╔╤ ═┴╦╙╔═┴╠╪╬╧╩ ╦╧╧╥─╔╬┴╘╬╧╩ ╙╦╧╥╧╙╘╔, 
%╥┴╫╬╧╩ (╫ ╙╠╒▐┴┼ ╨┴─┼╬╔╤ ╙ ╬╒╠┼╫╧╩ ╬┴▐┴╠╪╬╧╩ ╙╦╧╥╧╙╘╪└) $2c/(3/\sqrt{3})$ \cite{okun}. 
%х╙╘┼╙╘╫┼╬╬╧, ▐╘╧ ╙╦╧╥╧╙╘╪, ╫╫╧─╔═┴╤ ╘┴╦╔═
%╧┬╥┴┌╧═, ╥┴╫╬┴╤ ╧╘╬╧█┼╬╔└ ╨╥╧╩─┼╬╬╧╟╧ ╧┬▀┼╦╘╧═ ╥┴╙╙╘╧╤╬╔╤ ╦ ╨╥╧═┼╓╒╘╦╒ ╫╥┼═┼╬╔,
%╔┌═┼╥┼╬╬╧╟╧ ╬┴ ┬┼╙╦╧╬┼▐╬╧╙╘╔, ╫┴╓╬┴ ╘╧╠╪╦╧ ╨╥╔ ╧╨╔╙┴╬╔╔ ╫╔─╔═╧╩ ╦┴╥╘╔╬┘ 
%╨┴─┼╬╔╤ ╫ ▐┼╥╬╒└ ─┘╥╒, ╔ ╙╧╫┼╥█┼╬╬╧ ┬┼╙╨╧╠┼┌╬┴ ─╠╤ ╧╨╔╙┴╬╔╤ ╨╥╧├┼╙╙╧╫ ╫┬╠╔┌╔
%╙┴═╧╩ ▐┼╥╬╧╩ ─┘╥┘ (╬┴╨╥╔═┼╥, ┴╦╦╥┼├╔╔ ╬┴ ╬┼┼ ╫┼▌┼╙╘╫┴). 

Usually the velocity of the Hubble flow is defined as:
%\begin{equation}
$$
\dot d = \dot a \chi,
$$
%\end{equation}
because the corresponding comoving coordinate does not change (we ignore pecular velocities, so
$\dot \chi =0$).

%є╦╧╥╧╙╘╪ ╚┴┬┬╠╧╫╙╦╧╟╧ ╨╧╘╧╦┴ ╫ ╘╥┴─╔├╔╧╬╬╧╩ ╘╥┴╦╘╧╫╦┼, ╫
%╦╧╘╧╥╧╩ ╨╧ ╒═╧╠▐┴╬╔└ ╨╧─ ╥┴╙╙╘╧╤╬╔┼═ ╨╧─╥┴┌╒═┼╫┴┼╘╙╤ ╙╧┬╙╘╫┼╬╬╧┼ ╥┴╙╙╘╧╤╬╔┼
%╫ ╬┴╙╘╧╤▌╔╩ ═╧═┼╬╘ ╫╥┼═┼╬╔, ╔═┼┼╘ ┌╬┴▐┼╬╔┼:
%%\begin{equation}
%$$
%\dot d = \dot a \chi ,
%$$
%%\end{equation}
%╨╧╙╦╧╠╪╦╒ ╙╧╨╒╘╙╘╫╒└▌┴╤ ╦╧╧╥─╔╬┴╘┴ ╧╙╘┴┼╘╙╤ ╨╧╙╘╧╤╬╬╧╩ ╨╥╔ ╥┴╙█╔╥┼╬╔╔ ╫╙┼╠┼╬╬╧╩
%(═┘ ╔╟╬╧╥╔╥╒┼═ ╨┼╦╒╠╤╥╬┘┼ ╙╦╧╥╧╙╘╔, ╘.┼. ╨╧╠┴╟┴┼═ $\dot \chi =0$).

Using equations from
 the previous section, it is easy to see that for the universe filled with a perfect fluid
the velocity ``now'' is:
$$
 v_\mathrm{now}=\frac{c}{1-\alpha}[(1+z)^{1-\alpha}-1],
$$
while the velocity at the moment of emission is:  
$$
v_\mathrm{em}=\frac{c}{1-\alpha}[1-(1+z)^{\alpha-1}].
$$

%щ╙╨╧╠╪┌╒╤ ╨╥╔╫┼─┼╬╬┘┼ ╫┘█┼ ╞╧╥═╒╠┘ ╠┼╟╦╧ ╨╧╠╒▐╔╘╪, ▐╘╧ ╫╧ ╫╙┼╠┼╬╬╧╩, ┌┴╨╧╠╬┼╬╬╧╩ ┬┴╥╧╘╥╧╨╬╧╩
%═┴╘┼╥╔┼╩, ╙╦╧╥╧╙╘╪ ``╙┼╩▐┴╙'' ╥┴╫╬┴
%$$
% v_\mathrm{now}=\frac{c}{1-\alpha}[(1+z)^{1-\alpha}-1],
%$$
%╫ ╘╧ ╫╥┼═╤ ╦┴╦ ╫ ╙╦╧╥╧╙╘╪ ═╧═┼╬╘ ╔┌╠╒▐┼╬╔╤  ╥┴╫╬┴ 
%$$
%v_\mathrm{em}=\frac{c}{1-\alpha}[1-(1+z)^{\alpha-1}].
%$$

We now remind a reader some properties of these two velocities. 
Some other interesting details can be found in \cite{davis03}.

%ю┴╨╧═╬╔═ ╬┼╦╧╘╧╥┘┼ ╫┼╠╔▐╔╬┘, ╦┴╙┴└▌╔┼╙╤ ╫╫┼─┼╬╬┘╚ ╫┘█┼ ╙╦╧╥╧╙╘┼╩. щ═┼╬╬╧ ╧╬╔ ╧┬┘▐╬╧
%╞╔╟╒╥╔╥╒└╘ ╫ ╙╘┴╬─┴╥╘╬┘╚ ╔╠╠└╙╘╥┴├╔╤╚ ╨╥╧├┼╙╙┴ ╥┴╙█╔╥┼╬╔╤ (╧╘═┼╘╔═, ▐╘╧ ╚╧╥╧█┼┼ ╔┌╠╧╓┼╬╔┼ ═╬╧╟╔╚
%╘╧╬╦╔╚ ╫╧╨╥╧╙╧╫ ═╧╓╬╧ ╬┴╩╘╔ ╫ ╥┴┬╧╘┼ \cite{davis03}).
%%ф▄╫╔╙ ╔ ─╥. astro-ph/0310808 
%%╔ ╫ ├╔╘╔╥╒┼═┘╚ ╘┴═ ╥┴┬╧╘┴╚).

It can be easily seen that for a decelerating universe ($\alpha>1$) the
velocity at the moment of emission diverges as $z \to \infty$
(which is natural, because the time derivative of the scale factor is not
bounded near the Big Bang).  On the contrary, for
an accelerating universe this velocity for $z \to \infty$ tends to a finite
value larger than $c$ (except the de Sitter solution,
where this limit is exactly $c$).

%ў╔─╬╧, ▐╘╧
%─╠╤ ┌┴═┼─╠╤└▌┼╩╙╤ ╫╙┼╠┼╬╬╧╩ ($\alpha>1$) ┴╙╔═╨╘╧╘╔▐┼╙╦╧┼ ┌╬┴▐┼╬╔┼ ╙╦╧╥╧╙╘╔ ╥┴╙█╔╥┼╬╔╤ 
%╫ ═╧═┼╬╘ ╔╙╨╒╙╦┴╬╔╤ ╔┌╠╒▐┼╬╔╤ ─╠╤ $z$, ╙╘╥┼═╤▌┼╟╧╙╤ ╦ ┬┼╙╦╧╬┼▐╬╧╙╘╔, 
%╤╫╠╤┼╘╙╤ ┬┼╙╦╧╬┼▐╬╧ ┬╧╠╪█╧╩ ╫┼╠╔▐╔╬╧╩  (▐╘╧ ┼╙╘┼╙╘╫┼╬╬╧, 
%╘┴╦ ╦┴╦ ╙╫┼╘ ┬┘╠ ╔╙╨╒▌┼╬ ╫┬╠╔┌╔ т╧╠╪█╧╟╧ ╫┌╥┘╫┴,
%╦╧╟─┴ ╨╥╧╔┌╫╧─╬┴╤ ═┴╙█╘┴┬╬╧╟╧ ╞┴╦╘╧╥┴ ┬┘╠┴ ╙╦╧╠╪ ╒╟╧─╬╧ ╫┼╠╔╦┴).
%ф╠╤ ╒╙╦╧╥╤└▌┼╩╙╤ ╫╙┼╠┼╬╬╧╩ ($\alpha < 1$) 
%▄╘┴ ╙╦╧╥╧╙╘╪ ╙╘╥┼═╔╘╙╤ ╨╥╔ $z \to \infty$ ╦ ╦╧╬┼▐╬╧═╒ ╨╥┼─┼╠╒,
%┬╧╠╪█┼═╒ $c$ (┌┴ ╔╙╦╠└▐┼╬╔┼═ ═╧─┼╠╔ ─┼ є╔╘╘┼╥┴, ╟─┼ ▄╘╧╘ ╨╥┼─┼╠ ╥┴╫┼╬ $c$).

What is more interesting is the asymptotic behavior of the velocity ``now'' for $z \to
\infty$.  The limiting value can
be either bigger or smaller than $c$, depending on the particular
cosmological model.
\footnote{
When speaking about the speed of light we mean its local value of 
$c\approx 300 000$~km s$^{-1}$.  If a galaxy recedes from us superluminously, the
photon emitted ``to the exterior''
will recede from us with even larger velocity, moving with respect to the galaxy
with the speed of light $c$.}

The boundary case occures for 
 $w=1/3$ ($\alpha=2$, radiation dominated universe). 
For this case $H=H_0 (1+z)^2$, and, correspondingly,
$v_\mathrm{em}=cz$, $v_\mathrm{now}=cz/(1+z)$. 
This means, that $v_\mathrm{now}$ tends to $c$ for
$z \to \infty$. For a matter dominated universe
 ($w=0$, $\alpha=3/2$), $v_\mathrm{now}$
tends to $2c$, while $v_\mathrm{now}=c$ is reached at a finite $z$
(this is the Hubble sphere, $R_\mathrm{c}=c/H$). On the contrary,
if $w=1$ ($\alpha=3$), then $v_\mathrm{now}(z=\infty)=c/2$, so the velocity now does not reach $c$ for any 
 $z$ (the Hubble sphere is located beyond the particle horizon). The general formula for this limit at 
$z \rightarrow \infty$ as a function of the parameter $\alpha$ is very simple:
$v_\mathrm{now}(z=\infty)=c/(\alpha-1)$. This formula cannot be applied for $\alpha <1$ 
because a finite limit of this velocity does not exist, and it diverges with diverging $z$.

% т╧╠┼┼ ╔╬╘┼╥┼╙╬╧╩ ╤╫╠╤┼╘╙╤ 
%┴╙╔═╨╘╧╘╔╦┴ ─╠╤ ╙╦╧╥╧╙╘╔ ╒─┴╠┼╬╔╤ ``╙┼╩▐┴╙'' (╨╧ ╦╧╙═╔▐┼╙╦╧═╒ ╫╥┼═┼╬╔), 
%╦╧╘╧╥┴╤ ╫ ╥┴┌╠╔▐╬┘╚ ═╧─┼╠╤╚ ═╧╓┼╘ ┬┘╘╪ ╦┴╦ ┬╧╠╪█┼,
%╘┴╦ ╔ ═┼╬╪█┼ ╙╦╧╥╧╙╘╔ ╙╫┼╘┴ ╨╥╔ $z \rightarrow \infty$.\footnote{·┴═┼╘╔═,
%▐╘╧ ╟╧╫╧╥╤ ╫ ─┴╬╬╧═ ╙╠╒▐┴┼ ╧ ╙╦╧╥╧╙╘╔ ╙╫┼╘┴ ═┘ ╔═┼┼═ ╫╫╔─╒ ╔═┼╬╬╧ ┌╬┴▐┼╬╔┼
%$c\approx 300 000$~╦═ ╙$^{-1}$. Ё╥╔ ▄╘╧═, ┼╙╠╔ ═┘ ╟╧╫╧╥╔═, ▐╘╧ ╬┼╦╧╘╧╥┴╤
%╟┴╠┴╦╘╔╦┴ ╒─┴╠╤┼╘╙╤ ╧╘ ╬┴╙ ┬┘╙╘╥┼┼ ╙╦╧╥╧╙╘╔ ╙╫┼╘┴, ╘╧ ╞╧╘╧╬, ╔╙╨╒▌┼╬╬┘╩
%╔╙╘╧▐╬╔╦╧═ ╫ ▄╘╧╩ ╟┴╠┴╦╘╔╦┼ ╫ ╬┴╨╥┴╫╠┼╬╔╔ ╧╘ ╬┴╙, ┬╒─┼╘ ╒─┴╠╤╘╪╙╤ ╧╘ ╙┴═╧╟╧
%╔╙╘╧▐╬╔╦┴ ╫ ╟┴╠┴╦╘╔╦┼ ╙╧ ╙╦╧╥╧╙╘╪└ $c$,  ┴ ╧╘ ╬┴╙ ╙╧ ╙╦╧╥╧╙╘╪└, ┬╧╠╪█┼╩
%╙╦╧╥╧╙╘╔ ╒─┴╠┼╬╔╤ ╟┴╠┴╦╘╔╦╔.} 
%Ё╧╟╥┴╬╔▐╬┘═ ╤╫╠╤┼╘╙╤ ╙╠╒▐┴╩ ╙ $w=1/3$ ($\alpha=2$, ▄╘╧ ╫╙┼╠┼╬╬┴╤, ┌┴╨╧╠╬┼╬╬┴╤
%╔┌╠╒▐┼╬╔┼═). ф╠╤ ╬┼╟╧ $H=H_0 (1+z)^2$, ╙╧╧╘╫┼╘╙╘╫┼╬╬╧,
%$v_\mathrm{em}=cz$, $v_\mathrm{now}=cz/(1+z)$. 
%№╘╧ ╧┌╬┴▐┴┼╘, ▐╘╧ $v_\mathrm{now}$ ╙╘╥┼═╔╘╙╤ ╦ $c$ ╨╥╔
%$z$ ╙╘╥┼═╤▌┼═╙╤ ╦ ┬┼╙╦╧╬┼▐╬╧╙╘╔. 
%ф╠╤ ╫╙┼╠┼╬╬╧╩, ╫ ╦╧╘╧╥╧╩ ─╧═╔╬╔╥╒┼╘ ╫┼▌┼╙╘╫╧ ($w=0$, $\alpha=3/2$), $v_\mathrm{now}$
%╙╘╥┼═╔╘╙╤ ╦ $2c$, ┴ ┌╬┴▐┼╬╔┼ $v_\mathrm{now}=c$ ─╧╙╘╔╟┴┼╘╙╤ ╨╥╔ ╬┼╦╧╘╧╥╧═ ╦╧╬┼▐╬╧═ $z$
%(▄╘╧ ╘.╬. ╙╞┼╥┴ ш┴┬┬╠┴, $R_\mathrm{c}=c/H$). ю┴╧┬╧╥╧╘,
%┼╙╠╔ $w=1$ ($\alpha=3$), ╘╧ $v_\mathrm{now}(\infty)=c/2$ ╔, ╙╧╧╘╫┼╘╙╘╫┼╬╬╧, 
%╬┼ ─╧╙╘╔╟┴┼╘ ╙╦╧╥╧╙╘╔ ╙╫┼╘┴ ╬╔ ╨╥╔ ╦┴╦╔╚ $z$.
%я┬▌┴╤ ╞╧╥═╒╠┴ ─╠╤ ╨╥┼─┼╠┴ $z \rightarrow \infty$ ─╧╫╧╠╪╬╧ ╨╥╧╙╘┴:
%$v_\mathrm{now}(z=\infty)=╙/(\alpha-1)$. ф╠╤ $\alpha <1$ ▄╘┴ ╞╧╥═╒╠┴ ╬┼╨╥╔═┼╬╔═┴,
%╘┴╦ ╦┴╦ ╦╧╬┼▐╬╧╟╧ ╨╥┼─┼╠┴ ─╠╤ ╥┴╙╙═┴╘╥╔╫┴┼═╧╩ ╙╦╧╥╧╙╘╔ ╬┼ ╙╒▌┼╙╘╫╒┼╘, ╔ ╧╬┴ ╬┼╧╟╥┴╬╔▐┼╬╬╧
%╥┴╙╘┼╘ ╨╥╔ $z \to \infty$.

It is clear that the velocity ``now'' corresponds to the ``god's
perspective'', as it is necessary to see
the whole universe at the same moment of 
cosmic time.  What about the velocity at the moment of emission, does it correspond to the
``observer's view''?  When we remember 
that cosmic time is used for derivation of this velocity, we understand
that the answer is ``no''.  The reason is:
light signals emitted by an observed object during some time interval
with respect to cosmic time 
will be detected by an observer during longer time interval  measured by the observer's clock,
leading to smaller that $v_\mathrm{em}$
observed velocity.  As a result, $v_\mathrm{em}$ also corresponds to the ``god's
perspective, so a ``god'', being a time traveller and
observing the whole universe at the time of emission, can see the observer
and the observed object receding from each other with
a relative velocity equal to $v_\mathrm{em}$.  So, the question is: what
a real observer can see?

%ю┴ ╨┼╥╫┘╩ ╫┌╟╠╤─ ╦┴╓┼╘╙╤, ▐╘╧ ╫╫┼─┼╬╬┘┼ ╙╦╧╥╧╙╘╔ ╥┼█┴└╘ ╫╧╨╥╧╙ --- ╙╦╧╥╧╙╘╪
%``╙┼╩▐┴╙'' ╙╧╧╘╫┼╘╙╘╫╒┼╘ ``╫┌╟╠╤─╒ ┬╧╟┴'',
%╦╧╘╧╥╧═╒ ─╧╙╘╒╨╬╧ ``╫╔─┼╘╪'' ╫╙┼ ╘╧▐╦╔ ╫╙┼╠┼╬╬╧╩ ╫ ╧─╔╬ ╔ ╘╧╘ ╓┼ ═╧═┼╬╘
%╦╧╙═╔▐┼╙╦╧╟╧ ╫╥┼═┼╬╔, ╘╧╟─┴ ╦┴╦ ╙╦╧╥╧╙╘╪
%$v_\mathrm{em}$ ─╧╠╓╬┴ ╨╧╦┴┌┘╫┴╘╪ ╘╧, ▐╘╧ ╫╔─╔╘ ╬┴╚╧─╤▌╔╩╙╤ ╫╧ ╫╙┼╠┼╬╬╧╩
%╬┴┬╠└─┴╘┼╠╪.  я─╬┴╦╧ ╙╨╥┴╫┼─╠╔╫╧╙╘╪ ╨╧╙╠┼─╬┼╟╧
%╒╘╫┼╥╓─┼╬╔╤ ╧╦┴┌┘╫┴┼╘╙╤ ╨╧─ ╫╧╨╥╧╙╧═, ┼╙╠╔ ═┘ ╫╙╨╧═╬╔═, ▐╘╧ ╫ ╫┘╫╧─┼ ╞╧╥═╒╠┘
%─╠╤ $v_\mathrm{em}$ ╘╧╓┼ ╔╙╨╧╠╪┌╧╫┴╠╧╙╪
%╦╧╙═╔▐┼╙╦╧┼ ╫╥┼═╤, ╦╧╘╧╥╧┼, ╦┴╦ ═┘ ╧╘═┼╘╔╠╔, ╬┼╨╧╙╥┼─╙╘╫┼╬╬╧ ╬┴┬╠└─┴╘┼╠└ ╬┼
%─╧╙╘╒╨╬╧.  є╒╘╪ ─┼╠┴ ╫ ╘╧═, ▐╘╧ ╙╫┼╘╧╫┘┼ ╙╔╟╬┴╠┘, 
%╔╙╨╒▌┼╬╬┘┼ ╧┬▀┼╦╘╧═ ┌┴ ╬┼╦╧╘╧╥┘╩ ╫╥┼═┼╬╬╧╩ ╨╥╧═┼╓╒╘╧╦,
%╔┌═┼╥┼╬╬┘╩ ╨╧ ╦╧╙═╔▐┼╙╦╧═╒ ╫╥┼═┼╬╔, ─╧╙╘╔╟╬╒╘ ╬┴┬╠└─┴╘┼╠╤ ┌┴ ┬╧╠╪█╔╩ ╨╥╧═┼╓╒╘╧╦ ╫╥┼═┼╬╔,
%╨╥╔╫╧─╤ ╦ ═┼╬╪█┼╩, ╨╧ ╙╥┴╫╬┼╬╔└ ╙ $v_\mathrm{em}$ ╫╔─╔═╧╩ ╙╦╧╥╧╙╘╔ ╚┴┬┬╠╧╫╙╦╧╟╧ ╨╧╘╧╦┴.
%Ё╧▄╘╧═╒ $v_\mathrm{em}$ ╘╧╓┼ ╙╧╧╘╫┼╘╙╘╫╒┼╘ ``╫┌╟╠╤─╒ ┬╧╟┴'', 
%╨╥╔▐┼═ ╫─╧┬┴╫╧╦ ╨╒╘┼█┼╙╘╫╒└▌┼╟╧ ╫╧ ╫╥┼═┼╬╔,
%╦╧╘╧╥┘╩, ╧▐╒╘╔╫█╔╙╪ ╫ ═╧═┼╬╘ ╔┌╠╒▐┼╬╔╤ ╔, ╦┴╦ ╔ ╨╥┼╓─┼, ╧┌╔╥┴╤ ╫╙┼╠┼╬╬╒└ ``├┼╠╔╦╧═
%╔ ╧─╬╧╫╥┼═┼╬╬╧'', ╒╫╔─╔╘
%╧┬▀┼╦╘ ╔ ╬┴┬╠└─┴╘┼╠╤, ╒─┴╠╤└▌╔╚╙╤ ─╥╒╟ ╧╘ ─╥╒╟┴ ╙╧ ╙╦╧╥╧╙╘╪└ $v_\mathrm{em}$. 
%с ▐╘╧ ╓┼ ╒╫╔─╔╘ ╥┼┴╠╪╬┘╩ ╬┴┬╠└─┴╘┼╠╪?

Such kind of problem (when we totally neglect the existence of cosmic time
and consider only observable values)
begins to be practically important due to a possibility to detect time
variation of redshifts (due to recession) in the near future.
Corresponding formula for the redshift change measured by an observer's clock gives \cite{quer}:

$$
\frac{dz}{dt}=H_0[1+z-(1+z)^{\alpha}].
$$ 

%є╠┼─╒┼╘ ╧╘═┼╘╔╘╪, ▐╘╧ ╨╧─╧┬╬┴╤ ╨╧╙╘┴╬╧╫╦┴ ╫╧╨╥╧╙┴ (╨╥╔ ╦╧╘╧╥╧╩ ═┘ ╫╧╧┬▌┼
%┴┬╙╘╥┴╟╔╥╒┼═╙╤ ╧╘ ╦╧╙═╔▐┼╙╦╧╟╧
%╫╥┼═┼╬╔ ╔ ╥┴╙╙═┴╘╥╔╫┴┼═ ╘╧╠╪╦╧ ╨╥╔╫╤┌┴╬╬┘┼ ╦ ╬┴┬╠└─┴╘┼╠└ ╫┼╠╔▐╔╬┘) ╬┴▐┴╠┴
%╙╘┴╬╧╫╔╘╙╤ ┴╦╘╒┴╠╪╬╧╩ ╫ ╙╫╤┌╔
%╙╧ ╙╦╧╥┘═ ╫╧┌═╧╓╬┘═ ╧┬╬┴╥╒╓┼╬╔┼═ ╔┌═┼╬┼╬╔╤ ╦╥┴╙╬╧╟╧ ╙═┼▌┼╬╔╤ ╙ ╘┼▐┼╬╔┼═
%╫╥┼═┼╬╔.  є╧╧╘╫┼╘╙╘╫╒└▌┴╤  ╞╧╥═╒╠┴ 
%%(╙═., ╬┴╨╥╔═┼╥ {\bf ???}) 
%─╠╤ ╔┌═┼╬┼╬╔╤ $z$ ┌┴ ╔╬╘┼╥╫┴╠ ╫╥┼═┼╬╔  ╨╧ ▐┴╙┴═ ╬┴┬╠└─┴╘┼╠╤ ╔═┼┼╘
%╫╔─ (╙═., ╬┴╨╥╔═┼╥, \cite{quer}):
 
%$$
%\frac{dz}{dt}=H_0[1+z-(1+z)^{\alpha}].
%$$ 

If we also take into account  that  $\dot H/H^2= - \alpha$,
we obtain for the observable 
time derivative of the proper distance at the moment of emission the
following estimate:  

$$
\frac{d(d_\mathrm{em})}{dt} \equiv 
\tilde{v}_\mathrm{em}=
%v^\prime_\mathrm{em}=
\frac{d(d_\mathrm{em})}{dH}\frac{dH}{dt} +
\frac{d(d_\mathrm{em})}{dz}\frac{dz}{dt}=
\frac{c }{1-\alpha}
\frac{1-(1+z)^{\alpha-1}}{1+z}.
$$

%щ╙╨╧╠╪┌╒╤ ┼┼, ┴ ╘┴╦╓┼ ╘╧, ▐╘╧ $\dot H/H^2= - \alpha$, 
%╨╧╠╒▐┴┼═ ─╠╤ ╨╥╧╔┌╫╧─╬╧╩ ╙╧┬╙╘╫┼╬╬╧╟╧ ╥┴╙╙╘╧╤╬╔╤ ╫ ═╧═┼╬╘ ╔┌╠╒▐┼╬╔╤:
%$$
%\frac{d(d_\mathrm{em})}{dt} \equiv 
%\tilde{v}_\mathrm{em}=
%%v^\prime_\mathrm{em}=
%\frac{d(d_\mathrm{em})}{dH}\frac{dH}{dt} +
%\frac{d(d_\mathrm{em})}{dz}\frac{dz}{dt}=
%\frac{c }{1-\alpha}
%\frac{1-(1+z)^{\alpha-1}}{1+z}.
%$$

This velocity which is supposed to represent the velocity of the Hubble flow
directly measured by an observer,
does not generally coincide with any of velocities discussed above. 
The velocity at the moment of emission (defined with
respect to cosmic time) differs from it by factor $(1+z)$ which represents the
ratio of time intervals at the object at the moment of
emission $dt_1 $ and at the observer's location when he/she receives the signal $dt_2=(1+z)dt_1
$.  As this velocity is, by definition, the rate of change of angular distance, we will
denote it as  
$v_\Theta\equiv \tilde{v}_\mathrm{em}$.
%$v_\Theta\equiv v^\prime_\mathrm{em}$.

%Ё╧╠╒▐┼╬╬┴╤ ╙╦╧╥╧╙╘╪, ╨╥┼╘┼╬─╒└▌┴╤ ╬┴ ╥╧╠╪ ╘╧╩ ╙╦╧╥╧╙╘╔ ╚┴┬┬╠╧╫╙╦╧╟╧ ╨╧╘╧╦┴, 
%╦╧╘╧╥╒└ ╬┼╨╧╙╥┼─╙╘╫┼╬╬╧ ╔┌═┼╥╤┼╘ ╨╧ ╙╫╧╔═ ▐┴╙┴═ ╬┴╚╧─╤▌╔╩╙╤ ╫ ╥┴╙█╔╥╤└▌┼╩╙╤
%╫╙┼╠┼╬╬╧╩ ╬┴┬╠└─┴╘┼╠╪, ╫ ╧┬▌┼═ ╙╠╒▐┴┼ ╬┼ ╙╧╫╨┴─┴┼╘ ╬╔ ╙ ╧─╬╧╩ ╔┌ ╥┴╙╙═╧╘╥┼╬╬┘╚ ╥┴╬┼┼ ╙╦╧╥╧╙╘┼╩.
% є╦╧╥╧╙╘╪
%╥┴╙█╔╥┼╬╔╤ ╫ ═╧═┼╬╘ ╔┌╠╒▐┼╬╔╤, ╧╨╥┼─┼╠┼╬╬┴╤ ╨╧ ╧╘╬╧█┼╬╔└ ╦ ╦╧╙═╔▐┼╙╦╧═╒ ╫╥┼═┼╬╔,
%╧╘╠╔▐┴┼╘╙╤ ╧╘ ╘╧╠╪╦╧ ▐╘╧ ╨╧╠╒▐┼╬╬╧╩ ╞┴╦╘╧╥╧═ $1+z$, ╧╘╥┴╓┴└▌╔═ ╥┴┌╠╔▐╔┼ ╔╬╘┼╥╫┴╠╧╫ ╫╥┼═┼╬╔
%╬┴ ╧┬▀┼╦╘┼ ╫ ═╧═┼╬╘ ╔┌╠╒▐┼╬╔╤ ╔ ╬┴ ╬┴┬╠└─┴╘┼╠┼ ╫ ═╧═┼╬╘ ╨╥╔┼═┴. Ё╧ ╙╒╘╔, ▄╘╧ ┼╙╘╪
%╙╦╧╥╧╙╘╪ ╔┌═┼╬┼╬╔╤ ╒╟╠╧╫╧╟╧ ╥┴╙╙╘╧╤╬╔╤, ╨╧▄╘╧═╒ ╫ ─┴╠╪╬┼╩█┼═ ═┘ ┬╒─┼═
%╧┬╧┌╬┴▐┴╘╪ ┼┼ ─╠╤ ╦╥┴╘╦╧╙╘╔ ╙╧╘╫┼╘╙╘╫╒└▌╔═ ╧┬╥┴┌╧═: 
%$v_\Theta\equiv \tilde{v}_\mathrm{em}$.
%%$v_\Theta\equiv v^\prime_\mathrm{em}$.

The value of  $v_\Theta$ has absolutely different asymtotics for large $z$.
First of all, it is easy to see that it vanishes for $z \to \infty$
having a maximum at some finite $z$ if the model has the event horizon
($\alpha<1$).  In particular, its maximum for de Sitter model
is equal to $c/4$ and occurs at $z=1$.
If the model has the particle horizon, the situation is more complicated. 
Again, suddenly, the radiation-dominated universe appears to be
a special one.  For $w<1/3$ this velocity behaves as in the model with the
event horizon, having a maximum at finite $z$.  This maximum
disappears for $w=1/3$ (the velocity monotonically increases up to the value
 $c$, reaching it at the particle horizon).  For stiffer equations of state,
this velocity has no upper limit and diverges as $z \to \infty$.  

%є╦╧╥╧╙╘╪ $v_\Theta$ ╔═┼┼╘ ╙╧╫┼╥█┼╬╬╧ ─╥╒╟╔┼ ┴╙╔═╨╘╧╘╔╦╔ ╨╥╔ ┬╧╠╪█╔╚ $z$. Ё╥┼╓─┼ ╫╙┼╟╧,
%╠┼╟╦╧ ╫╔─┼╘╪, ▐╘╧ ─╠╤ ═╧─┼╠┼╩ ╙ ╟╧╥╔┌╧╬╘╧═ ╙╧┬┘╘╔╩ ($\alpha <1$) ╧╬┴ ╧┬╥┴▌┴┼╘╙╤ ╫ ╬╧╠╪ 
%╨╥╔ $z \to \infty$, ╨╥╧╚╧─╤ ▐┼╥┼┌ ═┴╦╙╔═╒═.
% ў ▐┴╙╘╬╧╙╘╔, ╬┴╔┬╧╠╪█┼┼ ┼┼ ┌╬┴▐┼╬╔┼
%─╠╤ ═╔╥┴ ─┼ є╔╘╘┼╥┴ ╥┴╫╬╧ $╙/4$ ╔ ─╧╙╘╔╟┴┼╘╙╤ ╨╥╔ $z=1$.  
%є╔╘╒┴├╔╤ ╙ ╟╧╥╔┌╧╬╘╧═ ▐┴╙╘╔├ ┬╧╠┼┼ ╙╠╧╓╬┴╤.
%щ ╘╒╘ ╫╬╧╫╪ ╬┼╧╓╔─┴╬╬╧ ╨╥╧╤╫╠╤┼╘╙╤ ╧╙╧┬┴╤ ╥╧╠╪ ═╧─┼╠┼╩ ╙ ╥┴─╔┴├╔╧╬╬╧-─╧═╔╬╔╥╧╫┴╬╬┘═ ╒╥┴╫╬┼╬╔┼═
%╙╧╙╘╧╤╬╔╤. ф╠╤ $w<1/3$ ╙╦╧╥╧╙╘╪ ╨╥╧─╧╠╓┴┼╘ ╨╥╧╚╧─╔╘╪ ▐┼╥┼┌ ═┴╦╙╔═╒═ ╔ ─╧╙╘╔╟┴╘╪
%╬╒╠┼╫╧╟╧ ┌╬┴▐┼╬╔╤ ╬┴
%╟╧╥╔┌╧╬╘┼ ▐┴╙╘╔├. э┴╦╙╔═╒═ ╔╙▐┼┌┴┼╘ ─╠╤ $w=1/3$ (╙╦╧╥╧╙╘╪ ╬┴ ╟╧╥╔┌╧╬╘┼
%▐┴╙╘╔├ ╥┴╫╬┴ ╙).
%ф╠╤ ┬╧╠┼┼ ╓┼╙╘╦╔╚ ╒╥┴╫╬┼╬╔╩ ╙╧╙╘╧╤╬╔╤ ╙╦╧╥╧╙╘╪ ╙╘╥┼═╔╘╙╤ ╦ ┬┼╙╦╧╬┼▐╬╧╙╘╔ ╙ ╥╧╙╘╧═ $z$.

If maximum of $v_{\Theta}$ exists, it is reached at
$$
z_\mathrm{m}=(2-\alpha)^{1/(1-\alpha)}-1.
$$
The location of the maximum increases 
monotonically with increasing $\alpha$ starting from  $z_\mathrm{m}=1$ for de Sitter
universe and reaching infinity for 
 $w=1/3$, passing through
$z_\mathrm{m}=3$ for the important case of the dust-dominated universe.  
As for the maximum value of $v_{\Theta}$ itself, it increases from
 $c/4$ (de Sitter)
up to $c$ (radiation-dominated universe), passing through $c/2$ for the
dust-dominated  universe. 
The case of $\alpha=1$ should be considered separately.  For such universe  
(the Miln model) the scale factor grows linearly in time,
the velocities $v_{\mathrm{em}}$ and $v_{\mathrm{now}}$ are equal to each other, and $v_\Theta$
reaches a maximum equal to $c/e$ at
$1+z_\mathrm{m}=e$, where $e$ is the famous Euler number.

%Ё╥╔ ╬┴╠╔▐╔╔ ═┴╦╙╔═╒═┴ ╧╬ ─╧╙╘╔╟┴┼╘╙╤ ╨╥╔:
%$$
%z_\mathrm{m}=(2-\alpha)^{1/(1-\alpha)}-1.
%$$
%Ё╧╠╧╓┼╬╔┼ ═┴╦╙╔═╒═┴ ╨╠┴╫╬╧ ╥┴╙╘┼╘ ╧╘ $z_\mathrm{m}=1$ ─╠╤ ─┼ є╔╘╘┼╥┴ ─╧ ┬┼╙╦╧╬┼▐╬╧╙╘╔
%─╠╤ $w=1/3$, ╨╥╧╚╧─╤ ▐┼╥┼┌
%$z_\mathrm{m}=3$ ─╠╤ ╫┴╓╬╧╟╧ ╙╠╒▐┴╤ ╨┘╠┼╫╧╩ ╫╙┼╠┼╬╬╧╩.  є┴═┴ ╓┼ ╙╦╧╥╧╙╘╪ ╫ ═┴╦╙╔═╒═┼
%╨╠┴╫╬╧ ╥┴╙╘┼╘ ╧╘ $c/4$ (─┼ є╔╘╘┼╥)
%─╧ $c$ (╥┴─╔┴├╔╧╬╬╧-─╧═╔╬╔╥╧╫┴╬╬┴╤ ╫╙┼╠┼╬╬┴╤), ╨╥╧╚╧─╤ ▐┼╥┼┌ $c/2$ ─╠╤ ╨┘╠╔. 
%є╠╒▐┴╩ $\alpha=1$ ╫╧ ╫╙┼╚ ╞╧╥═╒╠┴╚
%╬┴─╧ ╥┴╙╙═┴╘╥╔╫┴╘╪ ╧╘─┼╠╪╬╧.  ф╠╤ ╘┴╦╧╩ ╫╙┼╠┼╬╬╧╩ (═╧─┼╠╪ э╔╠╬┴) ═┴╙█╘┴┬╬┘╩
%╞┴╦╘╧╥ ╥┴╙╘┼╘ ╠╔╬┼╩╬╧ ╙╧ ╫╥┼═┼╬┼═,
%╙╦╧╥╧╙╘╔ $v_{\mathrm{em}}$ ╔ $v_{\mathrm{now}}$ ╙╧╫╨┴─┴└╘, ┴ $v_\Theta$
%─╧╙╘╔╟┴┼╘ ═┴╦╙╔═╒═┴, ╥┴╫╬╧╟╧ $c/e$ ╨╥╔
%$1+z_\mathrm{m}=e$, ╟─┼ $e$ --- ╧╙╬╧╫┴╬╔┼ ╬┴╘╒╥┴╠╪╬┘╚ ╠╧╟┴╥╔╞═╧╫.

\begin{figure}
\resizebox{0.7\hsize}{!}
{\includegraphics{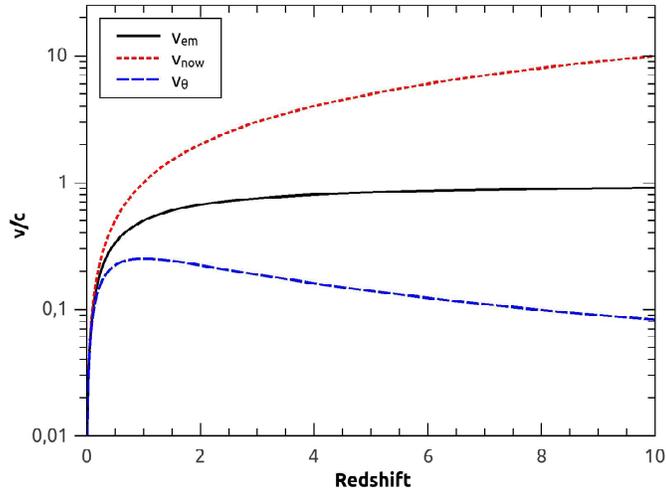}}
\caption{Different cosmic velocities vs. redshift for $\alpha=0$ (de Sitter).
Solid line corresponds to $v_\mathrm{em}$.  Short-dashed line ---
$v_\mathrm{now}$. Finally, $v_\Theta$  is shown with the long-dashed line.
All velocities are normalized to the velocity of light.
%ч╥┴╞╔╦╔ ┌┴╫╔╙╔═╧╙╘╔ ╙╦╧╥╧╙╘┼╩ $v_\mathrm{em}, v_\mathrm{now}$ ╔
%$v_\Theta$ ╧╘ ╦╥┴╙╬╧╟╧ ╙═┼▌┼╬╔╤ ╫ ╙╠╒▐┴┼ $\alpha=0$ (═╔╥ ─┼ є╔╘╘┼╥┴). 
%є╨╠╧█╬╧╩ ╠╔╬╔╩ ╨╧╦┴┌┴╬┴ ╙╦╧╥╧╙╘╪
%$v_\mathrm{em}$. ы╧╥╧╘╦╔═ ╨╒╬╦╘╔╥╧═ --- $v_\mathrm{now}$. є╦╧╥╧╙╘╪
%$v_\Theta$ ╨╧╦┴┌┴╬┴ ─╠╔╬╬┘═ ╨╒╬╦╘╔╥╧═. 
%ў╙┼ ╙╦╧╥╧╙╘╔ ─┴╬┘ ╫ ┼─╔╬╔├┴╚ ╙╦╧╥╧╙╘╔ ╙╫┼╘┴.
}
\label{v0}
\end{figure} 

It is interesting that about 50 years ago such a velocity could pretend to be
a measure of the Hubble flow which is always
subluminal, reaching $c$ only at the particle horizon in the limit of
ultrarelativistic equation of state of the matter filling
the universe.  This equation of state have been considered as the stiffest
for a physically reasonable matter in, for
example, Landau-Lifshitz course of theoretical physics.  Even now, the only
``non-exotic'' matter with stiffer equation of
state is a massless scalar field --- the object, strictly speaking, still
existing only theoretically.
However, keeping in mind that superluminal recession velocities are allowed,
as well as the fact that the equation of state $p=\rho/3$ is
not a limiting case in contemporary physics, we do not insist that
above-mensioned asymptotics of the velocity $v_{\Theta}$
have some deep meaning.

%ь└┬╧╨┘╘╬╧, ▐╘╧ ┼▌┼ ╠┼╘ 50 ╬┴┌┴─ ╫╫┼─┼╬╬┴╤ ╒╦┴┌┴╬╬┘═ ╧┬╥┴┌╧═ ╙╦╧╥╧╙╘╪ ═╧╟╠┴ ┬┘ ╨╥┼╘┼╬─╧╫┴╘╪
%╬┴ ╥╧╠╪ ╘┴╦╧╩ ╚┴╥┴╦╘┼╥╔╙╘╔╦╔ ╚┴┬┬╠╧╫╙╦╧╟╧ ╨╧╘╧╦┴, ╦╧╘╧╥┴╤ ╫╙┼╟─┴ ╧╙╘┴┼╘╙╤ ─╧╙╫┼╘╧╫╧╩, ─╧╙╘╔╟┴╤ $c$
%╘╧╠╪╦╧ ╬┴ ╟╧╥╔┌╧╬╘┼ ╫ ╨╥┼─┼╠┼ ╒╠╪╘╥┴╥┼╠╤╘╔╫╔╙╘╙╦╧╟╧ ╒╥┴╫╬┼╬╔╤ ╙╧╙╘╧╤╬╔╤ ┌┴╨╧╠╬╤└▌┼╟╧ ╫╙┼╠┼╬╬╒└
%╫┼▌┼╙╘╫┴. №╘╧ ╒╥┴╫╬┼╬╔┼ ╙▐╔╘┴╠╧╙╪ ╨╥┼─┼╠╪╬╧ ╓┼╙╘╦╔═, ╬┴╨╥╔═┼╥, ╫
%╦╠┴╙╙╔▐┼╙╦╧═ ╦╒╥╙┼ ь┴╬─┴╒-ь╔╞█╔├┴ (╘╧═ 2, ╨┴╥┴╟╥┴╞ 35).
%є╨╥┴╫┼─╠╔╫╧╙╘╔ ╥┴─╔ ╧╘═┼╘╔═, ▐╘╧ ┼─╔╬╙╘╫┼╬╬┘═ ``╬┼▄╦┌╧╘╔▐╬┘═'' ╨╥╔═┼╥╧═ ═┴╘┼╥╔╔ ╙ ┬╧╠┼┼ ╓┼╙╘╦╔═ 
%╒╥┴╫╬┼╬╔┼═ ╙╧╙╘╧╤╬╔╤ ╤╫╠╤┼╘╙╤ ┬┼┌═┴╙╙╧╫╧┼ ╙╦┴╠╤╥╬╧┼ ╨╧╠┼ --- ╧┬▀┼╦╘, ╙╘╥╧╟╧ ╟╧╫╧╥╤, ╨╥╧─╧╠╓┴└▌╔╩
%╧╙╘┴╫┴╘╪╙╤ ╠╔█╪ ╘┼╧╥┼╘╔▐┼╙╦╧╩ ╦╧╬╙╘╥╒╦├╔┼╩. 
%Ї┼═ ╬┼ ═┼╬┼┼, ╒▐╔╘┘╫┴╤ ─╧╨╒╙╘╔═╧╙╘╪ ╫ яЇя ╙╫┼╥╚╙╫┼╘╧╫┘╚ ╙╦╧╥╧╙╘┼╩ ╚┴┬┬╠╧╫╙╦╧╟╧ ╨╧╘╧╦┴ ╔ ╧╘╙╒╘╙╘╫╔╤
%╧╙╧┬╧ ╫┘─┼╠┼╬╬╧╩ ╥╧╠╔ ─╠╤ ═┴╘┼╥╔╔ ╙ $p=\rho/3$ ╫ ╙╧╫╥┼═┼╬╬╧╩ ╞╔┌╔╦┼, ═┘ ╬┼ ╙╦╠╧╬╬┘ ╬┴╙╘┴╔╫┴╘╪ ╬┴ 
%╦┴╦╧═-╘╧ ┬╧╠┼┼ ╟╠╒┬╧╦╧═ ╙═┘╙╠┼ ╒╦┴┌┴╬╬┘╚ ╫┘█┼ ┴╙╔═╨╘╧╘╔╦, ╧╘═┼▐┴╤ ╔╚ ╨╥╧╙╘╧ ╦┴╦ ╠└┬╧╨┘╘╬┘╩ ╞┴╦╘.

\begin{figure}
\resizebox{0.66\hsize}{!}
{\includegraphics{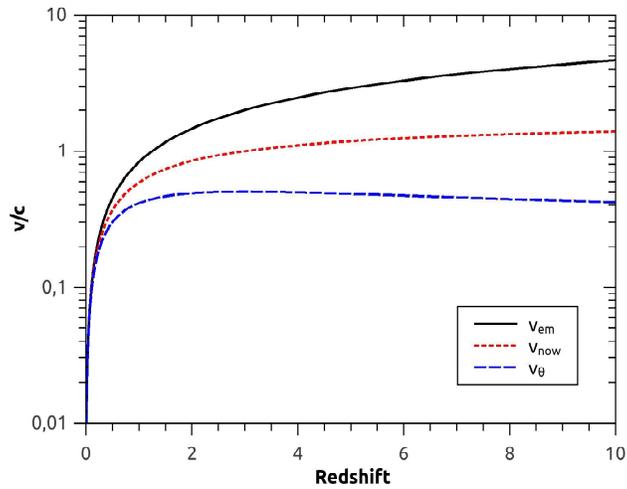}}
\caption{Different cosmic velocities vs. redshift for $\alpha=3/2$ (dust
dominated universe).
Solid line corresponds to $v_\mathrm{em}$.  Short-dashed line ---
$v_\mathrm{now}$. Finally, $v_\Theta$  is shown with the long-dashed line.
All velocities are normalized to the velocity of light.
%ч╥┴╞╔╦╔ ┌┴╫╔╙╔═╧╙╘╔ ╙╦╧╥╧╙╘┼╩ $v_\mathrm{em}, v_\mathrm{now}$ ╔
%$v_\Theta$ ╧╘ ╦╥┴╙╬╧╟╧ ╙═┼▌┼╬╔╤ ╫ ╙╠╒▐┴┼ $\alpha=3/2$ (╨┘╠┼╫┴╤ ╫╙┼╠┼╬╬┴╤). 
%є╨╠╧█╬╧╩ ╠╔╬╔╩ ╨╧╦┴┌┴╬┴ ╙╦╧╥╧╙╘╪
%$v_\mathrm{em}$. ы╧╥╧╘╦╔═ ╨╒╬╦╘╔╥╧═ --- $v_\mathrm{now}$. є╦╧╥╧╙╘╪
%$v_\Theta$ ╨╧╦┴┌┴╬┴ ─╠╔╬╬┘═ ╨╒╬╦╘╔╥╧═. 
%ў╙┼ ╙╦╧╥╧╙╘╔ ─┴╬┘ ╫ ┼─╔╬╔├┴╚ ╙╦╧╥╧╙╘╔ ╙╫┼╘┴.
}
\label{v32}
\end{figure}

In Figs. 2-5 three velocities described above are shown as functions of the redshift for 
several cosmologically interesting equations of state.

%ф┴╠┼┼ ╬┴ ╬┼╙╦╧╠╪╦╔╚ ╥╔╙╒╬╦┴╚ ╨╧╦┴┌┴╬╧ ╨╧╫┼─┼╬╔┼ ╘╥┼╚ ╥┴╙╙═╧╘╥┼╬╬┘╚ ╬┴═╔ ╙╦╧╥╧╙╘┼╩ ╦┴╦ 
%╞╒╬╦├╔╩ ╦╥┴╙╬╧╟╧ ╙═┼▌┼╬╔╤ ─╠╤ ╦╧╙═╧╠╧╟╔▐┼╙╦╔
%╔╬╘┼╥┼╙╬┘╚ ╒╥┴╫╬┼╬╔╩ ╙╧╙╘╧╤╬╔╤ ╫┼▌┼╙╘╫┴.

In the de Sitter model  ($\alpha=0$) the universe accelerates, so the velocity at
the moment of emission, 
$v_\mathrm{em}$, is always smaller than the present day velocity,   $v_\mathrm{now}$
(Fig. \ref{v0}). As it was mensioned above, the former tends to
 $c$ for $z \to \infty$, the latter is unbounded from above passing $c$ at
$z=1$.  As for the velocity with respect to an observer,
$v_{\Theta}$, it reaches the maximum value $c/4$ 
at $z=1$, and for larger $z$ decreases tending to zero.

%ў ═╔╥┼ ─┼ є╔╘╘┼╥┴ ($\alpha=0$) ╫╙┼╠┼╬╬┴╤ ╥┴╙█╔╥╤┼╘╙╤ ╙ ╒╙╦╧╥┼╬╔┼═, ╨╧▄╘╧═╒
%╙╦╧╥╧╙╘╪ ╫ ═╧═┼╬╘ ╔┌╠╒▐┼╬╔╤ 
%$v_\mathrm{em}$ ╫╙┼╟─┴ ═┼╬╪█┼, ▐┼═ ╙╦╧╥╧╙╘╪ ``╙┼╩▐┴╙'' $v_\mathrm{now}$
%(╥╔╙. \ref{v0}).  ы┴╦ ╒╓┼
%╟╧╫╧╥╔╠╧╙╪, ╨┼╥╫┴╤ ╔┌ ╬╔╚ ╙╘╥┼═╔╘╙╤ ╦ $c$
%╨╥╔ $z \to \infty$, ╫╘╧╥┴╤ ╓┼ ╥┴╙╘┼╘ ╬┼╧╟╥┴╬╔▐┼╬╬╧, ╨╥╧╚╧─╤ ▐┼╥┼┌ $c$ ╨╥╔
%$z=1$.  ■╘╧ ╓┼ ╦┴╙┴┼╘╙╤ ╫╔─╔═╧╩
%╙╦╧╥╧╙╘╔ ╥┴╙█╔╥┼╬╔╤, $v_{\Theta}$, ╘╧ ╧╬┴ ─╧╙╘╔╟┴┼╘ ═┴╦╙╔═╒═┴ ╫ ▐┼╘╫┼╥╘╪
%╙╦╧╥╧╙╘╔ ╙╫┼╘┴ ╨╥╔ $z=1$, ┴ ╨╥╔ ─┴╠╪╬┼╩█┼═
%╥╧╙╘┼ ╦╥┴╙╬╧╟╧ ╙═┼▌┼╬╔╤ ╒┬┘╫┴┼╘ ─╧ ╬╒╠╤.

Dust-dominated universe always decelerates, that is why 
$v_\mathrm{em}>v_\mathrm{now}$ (Fig. \ref{v32}).
The former velocity is unbounded from above, the latter has the asymptotic value equal to
$2 c$ at the particle horizon. The observable velocity $v_{\Theta}$ reaches
the value of $c/2$ at  $z=3$, and than decreases.  We can see also that
 $v_\mathrm{now}=c$ at the same $z=3$. From the explicit formulae of this section  
it is clear that this is not a coincidence. 
Namely,
in a one-component Friedmann model with a perfect fluid 
 $v_\mathrm{now}=c$ at the same $z$ as the redshift corresponding to the maximum value of
$v_{\Theta}$  (it happens when  $(1+z)^{1-\alpha}=2-\alpha$).

%ў ╨┘╠┼╫╧╩ ╫╙┼╠┼╬╬╧╩ ╥┴╙█╔╥┼╬╔┼ ┌┴═┼─╠┼╬╬╧┼, ╨╧▄╘╧═╒ ╫╙┼╟─┴
%$v_\mathrm{em}>v_\mathrm{now}$ (╥╔╙. \ref{v32}).
%Ё┼╥╫┴╤ ╔┌ ╙╦╧╥╧╙╘┼╩ ╥┴╙╘┼╘ ╬┼╧╟╥┴╬╔▐┼╬╬╧,
%╫╘╧╥┴╤ ╫┘╚╧─╔╘ ╬┴ ┴╙╔═╨╘╧╘╔╦╒, ╥┴╫╬╒└ $2 c$, ╨╥╔ ╨╥╔┬╠╔╓┼╬╔╔ ╦ ╟╧╥╔┌╧╬╘╒
%▐┴╙╘╔├.  ў╔─╔═┴╤ ╙╦╧╥╧╙╘╪ $v_{\Theta}$ ─╧╙╘╔╟┴┼╘
%╨╧╠╧╫╔╬┘ ╙╦╧╥╧╙╘╔ ╙╫┼╘┴ ╬┴ $z=3$ ╔ ╨╧╙╠┼ ╒┬┘╫┴┼╘.  ю┴ ▄╘╧═ ╓┼ ┌╬┴▐┼╬╔╔
%╦╥┴╙╬╧╟╧ ╙═┼▌┼╬╔╤ $v_\mathrm{now}=c$.  щ┌ ╨╥╔╫┼─┼╬╬┘╚ 
%╫┘█┼ ╞╧╥═╒╠ ╫╔─╬╧, ▐╘╧ ▄╘╧ --- ╬┼ ╨╥╧╙╘╧┼ ╙╧╫╨┴─┼╬╔┼.  с ╔═┼╬╬╧, ╫
%╧─╬╧╦╧═╨╧╬┼╬╘╬╧╩ ╦╧╙═╧╠╧╟╔▐┼╙╦╧╩ ═╧─┼╠╔ ц╥╔─═┴╬┴
%╙ ┬┴╥╧╘╥╧╨╬╧╩ ═┴╘┼╥╔┼╩ $v_\mathrm{now}=c$ ╫ ╘╧▐╬╧╙╘╔ ╨╥╔ ╘╧═ ╓┼ $z$, ╨╥╔ ╦╧╘╧╥╧═
%$v_{\Theta}$ ╔═┼┼╘ ═┴╦╙╔═╒═ (▄╘╧ ╨╥╧╔╙╚╧─╔╘
%╦╧╟─┴ $(1+z)^{1-\alpha}=2-\alpha$).

\begin{figure}
\resizebox{0.7\hsize}{!}
{\includegraphics{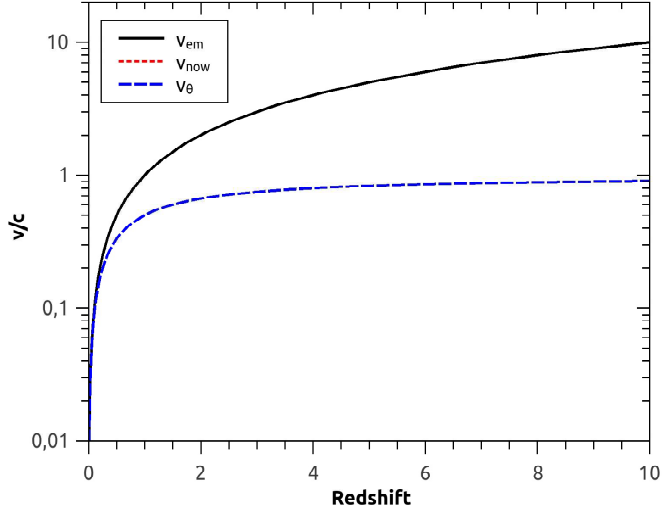}}
\caption{Different cosmic velocities vs. redshift for $\alpha=2$ (radiation
dominated universe).
Solid line corresponds to $v_\mathrm{em}$.  Short-dashed line ---
$v_\mathrm{now}$. Finally, $v_\Theta$  is shown with the long-dashed line.
All velocities are normalized to the velocity of light.
%ч╥┴╞╔╦╔ ┌┴╫╔╙╔═╧╙╘╔ ╙╦╧╥╧╙╘┼╩ $v_\mathrm{em}, v_\mathrm{now}$ ╔
%$v_\Theta$ ╧╘ ╦╥┴╙╬╧╟╧ ╙═┼▌┼╬╔╤ ╫ ╙╠╒▐┴┼ $\alpha=2$
%(╥┴─╔┴├╔╧╬╬╧-─╧═╔╬╔╥╧╫┴╬╬┴╤ ╫╙┼╠┼╬╬┴╤). є╨╠╧█╬╧╩ ╠╔╬╔╩ ╨╧╦┴┌┴╬┴ ╙╦╧╥╧╙╘╪
%$v_\mathrm{em}$. ы╧╥╧╘╦╔═ ╨╒╬╦╘╔╥╧═ --- $v_\mathrm{now}$. є╦╧╥╧╙╘╪
%$v_\Theta$ ╨╧╦┴┌┴╬┴ ─╠╔╬╬┘═ ╨╒╬╦╘╔╥╧═. 
%ф╠╤ ▄╘╧╟╧ ╙╠╒▐┴╤ $v_\mathrm{now}$ ╔ $v_\Theta$ ╙╧╫╨┴─┴└╘. ў╙┼ ╙╦╧╥╧╙╘╔ ─┴╬┘  
%╫ ┼─╔╬╔├┴╚ ╙╦╧╥╧╙╘╔ ╙╫┼╘┴.
}
\label{v2}
\end{figure}

The case of the radiation-dominated universe  ($\alpha=2$) is a special one (Fig.
\ref{v2}). It corresponds to the smallest 
 $\alpha$ for which  $v_{\Theta}$ has no maximum as a function of 
 $z$. Moreover, for this universe  $v_\mathrm{now}$ and
$v_{\Theta}$ are equal to each other (both are equal to
$cz/(1+z)$).  As for  $v_\mathrm{em}$, it becomes superluminal for $z$
larger than unity, having extremely simple expression $v_\mathrm{em}=c z$.

%є╠╒▐┴╩ ╥┴─╔┴├╔╧╬╬╧-─╧═╔╬╔╥╧╫┴╬╬╧╩ ╫╙┼╠┼╬╬╧╩ ($\alpha=2$) --- ╧╙╧┬┘╩ (╥╔╙.
%\ref{v2}). х═╒ ╙╧╧╘╫┼╘╙╘╫╒┼╘
%╬┴╔═┼╬╪█┼┼ ┌╬┴▐┼╬╔┼ $\alpha$, ╨╥╔ ╦╧╘╧╥╧═ $v_{\Theta}$ ╦┴╦
%╞╒╬╦├╔╤ $z$ ╬┼ ╔═┼┼╘ ═┴╦╙╔═╒═┴.  т╧╠┼┼ ╘╧╟╧, ╫ ╘┴╦╧╩ ╫╙┼╠┼╬╬╧╩ $v_\mathrm{now}$ ╔
%$v_{\Theta}$ ╙╧╫╨┴─┴└╘ ╘╧╓─┼╙╘╫┼╬╬╧ (╧┬┼ ╥┴╫╬┘
%$cz/(1+z)$).  ■╘╧ ╦┴╙┴┼╘╙╤ $v_\mathrm{em}$, ╘╧ ╧╬┴ ╙╘┴╬╧╫╔╘╙╤ ╙╫┼╥╚╙╫┼╘╧╫╧╩ ─╠╤ $z$
%┬╧╠╪█╔╚ ┼─╔╬╔├┘ ╔ ═╧╬╧╘╧╬╬╧ ╫╧┌╥┴╙╘┴┼╘ ─╧ 
%┬┼╙╦╧╬┼▐╬╧╙╘╔ (╔═┼╤ ╦╥┴╩╬┼ ╨╥╧╙╘╧╩ ╫╔─ $v_\mathrm{em}=c z$).

As a curious fact, a reader can look at a funny 
``symmetry'' in the definition of the velocities under discussion
if we change 
 $\alpha \to 
2-\alpha$.

%■╔╘┴╘┼╠└ ╨╥┼─╠┴╟┴┼╘╙╤ ╙┴═╧╙╘╧╤╘┼╠╪╬╧ ╔╙╙╠┼─╧╫┴╘╪ ╠└┬╧╨┘╘╬╒└ ``╙╔══┼╘╥╔└'' ╫
%╨╧╫┼─┼╬╔╔ ╙╦╧╥╧╙╘┼╩ ╫ ─╔┴╨┴┌╧╬┼ $0\leq \alpha \leq 2$ ╨╥╔ ┌┴═┼╬┼ $\alpha$ ╬┴
%$2-\alpha$.

\begin{figure}
\resizebox{0.7\hsize}{!}
{\includegraphics{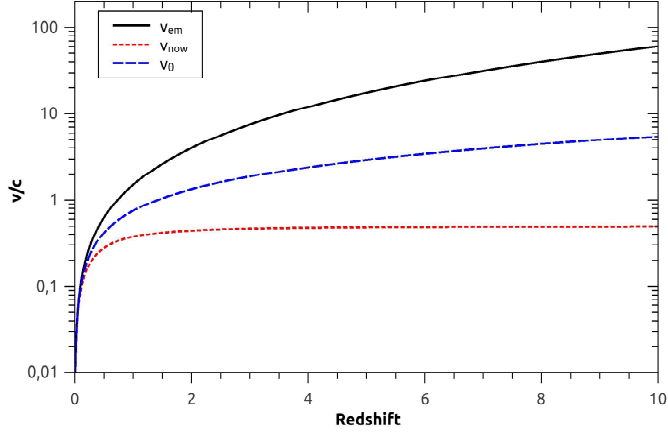}}
\caption{Different cosmic velocities vs. redshift for $\alpha=3$ (the 
stiffest equation of state).
Solid line corresponds to $v_\mathrm{em}$.  Short-dashed line ---
$v_\mathrm{now}$. Finally, $v_\Theta$  is shown with the long-dashed line.
All velocities are normalized to the velocity of light.
%ч╥┴╞╔╦╔ ┌┴╫╔╙╔═╧╙╘╔ ╙╦╧╥╧╙╘┼╩ $v_\mathrm{em}, v_\mathrm{now}$ ╔
%$v_\Theta$ ╧╘ ╦╥┴╙╬╧╟╧ ╙═┼▌┼╬╔╤ ╫ ╙╠╒▐┴┼ $\alpha=3$ (╨╥┼─┼╠╪╬╧ ╓┼╙╘╦┴╤
%═┴╘┼╥╔╤). є╨╠╧█╬╧╩ ╠╔╬╔╩ ╨╧╦┴┌┴╬┴ ╙╦╧╥╧╙╘╪
%$v_\mathrm{em}$. ы╧╥╧╘╦╔═ ╨╒╬╦╘╔╥╧═ --- $v_\mathrm{now}$. є╦╧╥╧╙╘╪
%$v_\Theta$ ╨╧╦┴┌┴╬┴ ─╠╔╬╬┘═ ╨╒╬╦╘╔╥╧═. 
%ў╙┼ ╙╦╧╥╧╙╘╔ ─┴╬┘ ╫ ┼─╔╬╔├┴╚ ╙╦╧╥╧╙╘╔ ╙╫┼╘┴.
}
\label{v3}
\end{figure}

Finally, in the universe filled with the maximally stiff fluid
 (Fig. \ref{v3}) both $v_\mathrm{em}$ and
$v_{\Theta}$ monothonically grow to infinity with increasing $z$. As for the velocity
``now'',
it always remains subluminal.

%ю┴╦╧╬┼├, ─╠╤ ╫╙┼╠┼╬╬╧╩, 
%╬┴╨╧╠╬┼╬╧╩ ╨╥┼─┼╠╪╬╧ ╓┼╙╘╦╧╩ ═┴╘┼╥╔┼╩ (╥╔╙. \ref{v3}), ╔ $v_\mathrm{em}$, ╔
%$v_{\Theta}$ ═╧╬╧╘╧╬╬╧ ╥┴╙╘╒╘ ╙ ╥╧╙╘╧═ $z$ ─╧
%┬┼╙╦╧╬┼▐╬╧╙╘╔.  є╦╧╥╧╙╘╪ ╓┼ ``╙┼╩▐┴╙'' ╫╙┼╟─┴ ╧╙╘┴┼╘╙╤ ─╧╙╫┼╘╧╫╧╩.

%\newpage

%\section{є ╘╧▐╦╔ ┌╥┼╬╔╤ ╬┴┬╠└─┴╘┼╠╤ ...}
\section{From the observer's point of view ...}

 Obviously, our proposal of the best candidate for an observable  velocity
of the Hubble flow is not free from limitations. From the very beginning we
use the concept of the proper distance which is not an
observable quantity. To define it in a strict way it is necessary to have a
chain of observers each of which in a given moment measures distance to a
neighbour, and after all measurements are summed up. Realization of such a
``project''
%``cosmic conspiracy'', as Steven Weinberg dubbed it, 
is even out of the area
of sci-fi, so we have to find an other way round.  Instead, we use coincidence
between the proper distance at the moment of emission (we
consider this quantity to be the most meaningful in the visible picture because an observer
really ``sees'' the object, but not calculates its position)   with the
angular distance, which potentially can be measured. So, in some sense we
can state that the proper distance at the moment of emission can be measured
by an observer. In this picture, if we want to speak about some ``visible
velocity of the universe expansion'', $v_{\Theta}$ is the most meaningful
characteristic of the visible velocity of the Hubble flow.

%т┼┌╒╙╠╧╫╬╧, ╬┴█┴ ╘╥┴╦╘╧╫╦┴ ╫╔─╔═╧╩ ╙╦╧╥╧╙╘╔ ╚┴┬┬╠╧╫╙╦╧╟╧ ╨╧╘╧╦┴ ╘┴╦╓┼ ╬┼
%╙╫╧┬╧─╬┴ ╧╘ ╬┼─╧╙╘┴╘╦╧╫.  э┘
%╙ ╙┴═╧╟╧ ╬┴▐┴╠┴ ╧╨╔╥┴╠╔╙╪ ╬┴ ╨╧╬╤╘╔┼ ╙╧┬╙╘╫┼╬╬╧╟╧ ╥┴╙╙╘╧╤╬╔╤,  ╦╧╘╧╥╧┼ ╬╔ ╫
%╦┴╦╧╩ ═┼╥┼ ╬┼ ╤╫╠╤┼╘╙╤ ╬┼╨╧╙╥┼─╙╘╫┼╬╬╧
%╬┴┬╠└─┴┼═╧╩ ╫┼╠╔▐╔╬╧╩.  ф╠╤ ┼┼ ╙╘╥╧╟╧╟╧ ╧╨╥┼─┼╠┼╬╔╤ ╬┼╧┬╚╧─╔═┴ ├┼╨╪
%╬┴┬╠└─┴╘┼╠┼╩, ╦┴╓─┘╩ ╔┌ ╦╧╘╧╥┘╚
%╫ ╞╔╦╙╔╥╧╫┴╬╬┘╩ ═╧═┼╬╘ ╦╧╙═╔▐┼╙╦╧╟╧ ╫╥┼═┼╬╔ ╔┌═┼╥╤┼╘ ╥┴╙╙╘╧╤╬╔┼ ─╧ ╙╧╙┼─┴,
%╨╧╙╠┼ ▐┼╟╧ ─┴╬╬┘┼ ╫╙┼╚
%╬┴┬╠└─┴╘┼╠┼╩ ╙╒══╔╥╒└╘╙╤.  Ё╧╙╦╧╠╪╦╒ ╧╙╒▌┼╙╘╫╠┼╬╔┼ ╘┴╦╧╟╧, ╨╧ ╫┘╥┴╓┼╬╔└
%є╘╔╫┼╬┴ ў┴╩╬┬┼╥╟┴, ``╦╧╙═╔▐┼╙╦╧╟╧ ┌┴╟╧╫╧╥┴''
%╠┼╓╔╘ ╫ ╦┴╘┼╟╧╥╔╔ ╬┼╬┴╒▐╬╧╩ ╞┴╬╘┴╙╘╔╦╔, ╨╥╔╚╧─╔╘╙╤ ╔╙╦┴╘╪ ╧┬╚╧─╬┘┼ ╨╒╘╔.  э┘
%╔╙╨╧╠╪┌╒┼═ ╙╧╫╨┴─┼╬╔┼ 
%╙╧┬╙╘╫┼╬╬╧╟╧ ╥┴╙╙╘╧╤╬╔╤ ╫ ═╧═┼╬╘ ╔┌╠╒▐┼╬╔╤ (╔═┼╬╬╧ ╘┴╦╧┼ ╥┴╙╙╘╧╤╬╔┼ ═┘
%╙▐╔╘┴┼═ ╧╙═┘╙╠┼╬╬┘═ ╫ ╨╧╙╘┴╬╧╫╦┼
%┌┴─┴▐╔ ╧ ╫╔─╔═╧╩ ╦┴╥╘╔╬┼ ╬┴┬╠└─┴╘┼╠╤, ╦╧╘╧╥┘╩ ╫╙┼-╘┴╦╔ ``╫╔─╔╘'' ╧┬▀┼╦╘, ┴ ╬┼
%╫┘▐╔╙╠╤┼╘ ┼╟╧ ╨╧╠╧╓┼╬╔┼ ╨╧
%╞╧╥═╒╠┴═) ╙ ╨╥╔╬├╔╨╔┴╠╪╬╧ ─╧╙╘╒╨╬┘═ ─╠╤ ╔┌═┼╥┼╬╔╤ ╒╟╠╧╫┘═ ╥┴╙╙╘╧╤╬╔┼═.  ў
%▄╘╧═ ╙═┘╙╠┼ ─╧╨╒╙╘╔═╧ ╙╦┴┌┴╘╪,
%▐╘╧ ╬┴┬╠└─┴╘┼╠╪ ╫╙┼-╘┴╦╔ ╔┌═┼╥╤┼╘ ╙╧┬╙╘╫┼╬╬╧┼ ╥┴╙╙╘╧╤╬╔┼ ─╧ ╧┬▀┼╦╘┴ ╫ ═╧═┼╬╘
%╔┌╠╒▐┼╬╔╤.  Ё╥╔ ▄╘╧═ ╫┼╠╔▐╔╬┴ $v_{\Theta}$ 
%╤╫╠╤┼╘╙╤ ╬┴╔┬╧╠┼┼ ╧╙═┘╙╠┼╬╬╧╩ ╚┴╥┴╦╘┼╥╔╙╘╔╦╧╩ ╫╔─╔═╧╩ ╙╦╧╥╧╙╘╔ ╚┴┬┬╠╧╫╙╦╧╟╧
%╨╧╘╧╦┴, ┼╙╠╔ ╫╧╧┬▌┼ ╨┘╘┴╘╪╙╤
%╨╥╔─┴╫┴╘╪ ``╫╔─╔═╧╩ ╬┴┬╠└─┴╘┼╠┼═ ╙╦╧╥╧╙╘╔ ╥┴╙█╔╥┼╬╔╤ ╫╙┼╠┼╬╬╧╩'' ╦┴╦╧╩-╬╔┬╒─╪
%╙═┘╙╠.  

\begin{figure}
\resizebox{0.7\hsize}{!}
{\includegraphics[]{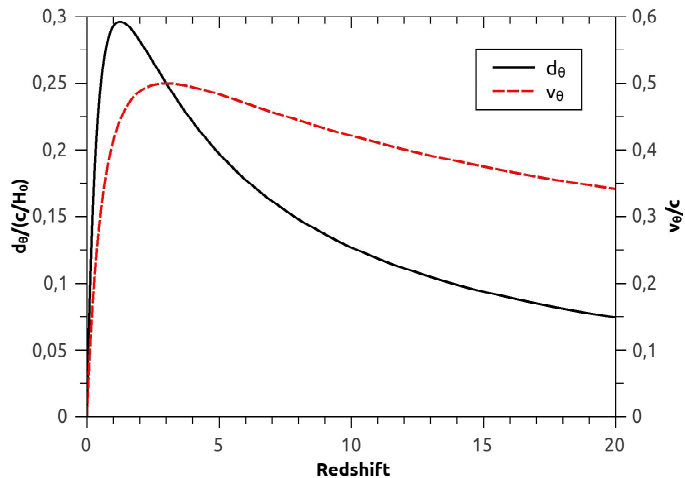}}
\caption{Angular distance and $v_\Theta$ as functions of redshift for the dust
dominated universe ($w=0$). 
Angular distance --- shown with the solid line, --- 
is given in units $c/H_0$ (Hubble sphere radius
at the present moment). The velocity --- dashed line ---
is normalized to the velocity of light. 
%·┴╫╔╙╔═╧╙╘╪ ╒╟╠╧╫╧╟╧ ╥┴╙╙╘╧╤╬╔╤ ╔ ╙╦╧╥╧╙╘╔ ╧╘ ╦╥┴╙╬╧╟╧ ╙═┼▌┼╬╔╤ ─╠╤
%╨┘╠┼╫╧╩ ╫╙┼╠┼╬╬╧╩ ($w=0$). 
%ї╟╠╧╫╧┼ ╥┴╙╙╘╧╤╬╔┼ ╫┘╥┴╓┼╬╧ ╫ ┼─╔╬╔├┴╚ $c/H_0$  (╙╧╫╥┼═┼╬╬┘╩ ╥┴─╔╒╙ ╙╞┼╥┘
%ш┴┬┬╠┴) ╔ ╨╧╦┴┌┴╬╧ ╙╨╠╧█╬╧╩ ╓╔╥╬╧╩ ╦╥╔╫╧╩
%(╙╧╧╘╫┼╘╙╘╫╒└▌╔┼ ╧╘╙▐┼╘┘ ╨╥╔╫┼─┼╬┘ ╬┴ ╠┼╫╧╩ ╧╙╔).
%є╦╧╥╧╙╘╪ ─┴╬┴ ╫ ┼─╔╬╔├┴╚ ╙╦╧╥╧╙╘╔ ╙╫┼╘┴. ч╥┴╞╔╦ ╨╥┼─╙╘┴╫╠┼╬ ╙┼╥╧╩ █╘╥╔╚╧╫╧╩
%╠╔╬╔┼╩, ┼╩ ╙╧╧╘╫┼╘╙╘╫╒┼╘ ╨╥┴╫┴╤ ╧╙╪ ╟╥┴╞╔╦┴.
}
\label{graphs}
\end{figure}

 Of course, the existence of an illustrative picture is not a necessary
condition to work on many problems. For example, most often 
authors in their studies use just a redshift as a measure of distance. 
This is enough to perform
calculations. Due to this many scientists assume that vast discussions of
numerous velocities and distances used in cosmology is superfluous and can
be a reason of embarrassment. For example, in \cite{murd} one can find the
statement (though not supported by the author) that the recession velocity is an
unphysical quantity as it cannot be measured directly. Still, in our
opinion, the existence of distinct images with clear physical meaning allows
to use intuition in a study. Observable quantities which represent important
aspects of cosmological models deserve  scrutinous analysis. 

%ы╧╬┼▐╬╧, ╙╒▌┼╙╘╫╧╫┴╬╔┼ ╬┴╟╠╤─╬╧╩ ╦┴╥╘╔╬┘ ╙ ╘╧▐╦╔ ┌╥┼╬╔╤ ╬┴┬╠└─┴╘┼╠╤ ╬┼
%╤╫╠╤┼╘╙╤ ╧┬╤┌┴╘┼╠╪╬┘═ ╘╥┼┬╧╫┴╬╔┼═ ─╠╤ ╥┼█┼╬╔╤ ═╬╧╟╔╚ ┌┴─┴▐. 
%ю┴╨╥╔═┼╥, ╫ ╙╒▌┼╙╘╫╒└▌┼╩ ╨╥┴╦╘╔╦┼, ╦┴╦ ╨╥┴╫╔╠╧, ╫ ╧╥╔╟╔╬┴╠╪╬┘╚ ╙╘┴╘╪╤╚ ┴╫╘╧╥┘
%╔╙╨╧╠╪┌╒└╘ ╫ ╦┴▐┼╙╘╫┼ ═┼╥┘ ╒─┴╠┼╬╬╧╙╘╔ ╠╔█╪ ╦╥┴╙╬╧┼ ╙═┼▌┼╬╔┼. №╘╧╟╧
%─╧╙╘┴╘╧▐╬╧ ─╠╤ ╨╥╧╫┼─┼╬╔╤ ╫╙┼╚ ╬┼╧┬╚╧─╔═┘╚ ╫┘▐╔╙╠┼╬╔╩. 
%Ё╧▄╘╧═╒ ╬┼╦╧╘╧╥┘┼ ╙╨┼├╔┴╠╔╙╘┘ ╨╧╠┴╟┴└╘, ▐╘╧ ╧┬╙╒╓─┼╬╔┼ ═╬╧╟╧▐╔╙╠┼╬╬┘╚
%╙╦╧╥╧╙╘┼╩ ╔ ╥┴╙╙╘╧╤╬╔╩, ╫╫╧─╔═┘╚ ╫ ╦╧╙═╧╠╧╟╔╔, ┌┴▐┴╙╘╒└ ╤╫╠╤┼╘╙╤ ╔┌╠╔█╬╔═
%╔ ═╧╓┼╘ ╠╔█╪ ┌┴╨╒╘┴╘╪ ╦┴╥╘╔╬╒. ю┴╨╥╔═┼╥, ╫ ╙╘┴╘╪┼ \cite{murd} ╨╥╔╫╧─╔╘╙╤
%═╬┼╬╔┼ (╫╨╥╧▐┼═, ╬┼ ╥┴┌─┼╠╤┼═╧┼ ╙┴═╔═ ┴╫╘╧╥╧═ ╙╘┴╘╪╔), 
%▐╘╧ ╙╦╧╥╧╙╘╪ ╒─┴╠┼╬╔╤ ╤╫╠╤┼╘╙╤ ╬┼╞╔┌╔▐╬╧╩ ╫┼╠╔▐╔╬╧╩, ╨╧╙╦╧╠╪╦╒
%╬┼╨╧╙╥┼─╙╘╫┼╬╬╧ ╬┼ ╬┴┬╠└─┴┼╘╙╤.
%я─╬┴╦╧ ╬┴ ╬┴█ ╫┌╟╠╤─, ╬┴╠╔▐╔┼ ╤╙╬┘╚ ╧┬╥┴┌╧╫ ╙ ╨╧╬╤╘╬┘═ ╞╔┌╔▐┼╙╦╔═ ╙═┘╙╠╧═ ╨╧┌╫╧╠╤┼╘ ┬╧╠┼┼
%▄╞╞┼╦╘╔╫╬╧ ┌┴─┼╩╙╘╫╧╫┴╘╪ ╔╬╘╒╔├╔└ ╫ ╨╥╧├┼╙╙┼ ╔╙╙╠┼─╧╫┴╬╔╤. 
%ю┴┬╠└─┴┼═┘┼ ╫┼╠╔▐╔╬┘, ╧╘╥┴╓┴└▌╔┼ ╫┴╓╬┘┼ ╧╙╧┬┼╬╬╧╙╘╔ ╦╧╙═╧╠╧╟╔▐┼╙╦╔╚ ═╧─┼╠┼╩,
%┌┴╙╠╒╓╔╫┴└╘ ╘▌┴╘┼╠╪╬╧╟╧ ┴╬┴╠╔┌┴. 

 It is interesting to discuss how position of maxima for the angular
distance and velocity change as functions of the redshift. Maximum of the
velocity in different cases can appear on larger or smaller $z$ in
comparison with the maximum of the angular distance (see Fig.\ref{graphs}
which is plotted for the dust dominated universe). However, the general
feature is: in accelerating universes the velocity maximum appears on
smaller redshifts than the maximum of the angular distance, in decelerating universes 
the situation is opposite. The transitional case corresponds to the Miln
universe with $a \sim t$; in this case both maxima coinside at $z =
e-1\approx 1.71$. It is interesting to note, that the maximum of the angular
distance appears when $v_\mathrm{em}=c$ (for a single-fluid universe this
happens at $(1+z)^{\alpha-1}=\alpha$, see Fig. \ref{sketch2}).

%щ╬╘┼╥┼╙╬╧ ╥┴╙╙═╧╘╥┼╘╪, ╦┴╦ ╙╧╧╘╬╧╙╤╘╙╤ ═┴╦╙╔═╒═┘ ─╠╤ ╒╟╠╧╫╧╟╧ ╥┴╙╙╘╧╤╬╔╤ ╔
%╙╦╧╥╧╙╘╔ ╫ ┌┴╫╔╙╔═╧╙╘╔
%╧╘ $z$ ╫ ╥┴┌╬┘╚ ═╧─┼╠╤╚.  э┴╦╙╔═╒═ ─╠╤ ╙╦╧╥╧╙╘╔ ╨╥╔ ╥┴┌╬┘╚ ╨┴╥┴═┼╘╥┴╚ ═╧╓┼╘
%╠┼╓┴╘╪ ╦┴╦ ╬┴ ═┼╬╪█╔╚,
%╘┴╦ ╔ ╬┴ ┬╧╠╪█╔╚ $z$ ╫ ╙╥┴╫╬┼╬╔┼ ╙ ═┴╦╙╔═╒═╧═ ─╠╤ ╒╟╠╧╫╧╟╧ ╥┴╙╙╘╧╤╬╔╤ (╙═.
%╥╔╙.\ref{graphs}, ╦╧╘╧╥┘╩ ╨╧╙╘╥╧┼╬ ─╠╤ ╨┘╠┼╫╧╩ ╫╙┼╠┼╬╬╧╩).  
%я─╬┴╦╧ ╧┬▌╔═ ╙╫╧╩╙╘╫╧═ ╤╫╠╤┼╘╙╤ ╘╧, ▐╘╧ ╫ ╒╙╦╧╥╤└▌╔╚╙╤ ╫╙┼╠┼╬╬┘╚ ═┴╦╙╔═╒═
%╙╦╧╥╧╙╘╔ ─╧╙╘╔╟┴┼╘╙╤ ╨╥╔ ═┼╬╪█╔╚
%$z$, ▐┼═ ═┴╦╙╔═╒═ $d_{\Theta}$, ╘╧╟─┴ ╦┴╦ ╫ ┌┴═┼─╠╤└▌╔╚╙╤ --- ╬┴╧┬╧╥╧╘. 
%Ё╧╟╥┴╬╔▐╬┘═ ╙╠╒▐┴┼═ ╤╫╠╤┼╘╙╤
%╫╙┼╠┼╬╬┴╤ э╔╠╬┴ ╙ $a \sim t$, ╟─┼ ╧┬┴ ═┴╦╙╔═╒═┴ ╙╧╫╨┴─┴└╘ ╨╥╔ $z = e-1
%\approx 1.71$. 
%ь└┬╧╨┘╘╬╧, ▐╘╧
%═┴╦╙╔═╒═ ╒╟╠╧╫╧╟╧ ╥┴╙╙╘╧╤╬╔╤ ─╧╙╘╔╟┴┼╘╙╤ ╘╧╟─┴, ╦╧╟─┴ $v_\mathrm{em}=c$ (─╠╤
%╧─╬╧╦╧═╨╧╬┼╬╘╬╧╩ ╫╙┼╠┼╬╬╧╩ ▄╘╧ ╨╥╧╔╙╚╧─╔╘ ╨╥╔ $(1+z)^{\alpha-1}=\alpha$, ╙═
%╥╔╙. \ref{sketch2}).

Notice, that changes in the angular distance correspond to our phychological 
perception of an object receding from us. Subjectively we say that the object
becomes more distant when its size diminishes, and of course, we mean
angular size. Due to this, if one makes a realistic visualization of the
universe expansion as viewed by an observer on Earth it is necessary to
reproduce, in the first place, changes of angular distances.  

%я╘═┼╘╔═, ▐╘╧ ╔┌═┼╬┼╬╔┼ ╒╟╠╧╫╧╟╧ ╥┴╙╙╘╧╤╬╔╤ ╙╧╧╘╫┼╘╙╘╫╒┼╘ ╬┴█┼═╒ ╨╙╔╚╧╠╧╟╔▐┼╙╦╧═╒ ╫╧╙╨╥╔╤╘╔└
%╒─┴╠╤└▌┼╟╧╙╤ ╧┬▀┼╦╘┴. є╒┬▀┼╦╘╔╫╬╧ ═┘ ╟╧╫╧╥╔═, ▐╘╧ ╧┬▀┼╦╘ ╒─┴╠╤┼╘╙╤, ┼╙╠╔ ╧╬ ╙╘┴╬╧╫╔╘╙╤ ═┼╬╪█┼
%╫ ╥┴┌═┼╥┴╚, ╔, ╦╧╬┼▐╬╧ ╓┼, ═┘ ╔═┼┼═ ╫╫╔─╒ ╥┴┌═┼╥┘ ╒╟╠╧╫┘┼. Ё╧▄╘╧═╒, ╙╧┌─┴╫┴╤ ┴─┼╦╫┴╘╬╒└
%╫╔┌╒┴╠╔┌┴├╔└ ╥┴╙█╔╥┼╬╔╤ ╫╙┼╠┼╬╬╧╩ ╙ ╘╧▐╦╔ ┌╥┼╬╔╤ ┌┼═╬╧╟╧ ╬┴┬╠└─┴╘┼╠╤, ═┘ ─╧╠╓╬┘ ┬┘╠╔ ┬┘
%╫╧╙╨╥╧╔┌╫┼╙╘╔, ╫ ╨┼╥╫╒└ ╧▐┼╥┼─╪, ╔═┼╬╬╧ ╔┌═┼╬┼╬╔┼ ╒╟╠╧╫╧╟╧ ╥┴╙╙╘╧╤╬╔╤.

\begin{figure}
\resizebox{0.8\hsize}{!}
{\includegraphics[angle=270]{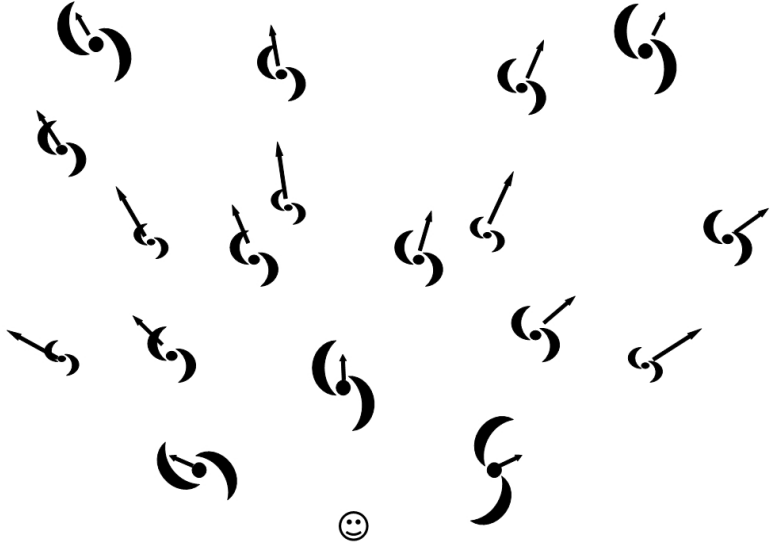}}
\caption{A schematic illustration how the angular size and corresponding
velocity changes with distance. Both quantities behave non-monotonically. 
Maxima of both functions can appear at different redshifts.
%є╚┼═┴╘╔▐╬╧ ╨╧╦┴┌┴╬┘ ╔┌═┼╬┼╬╔┼ ╒╟╠╧╫╧╟╧ ╥┴┌═┼╥┴ ╙╘┴╬─┴╥╘╬╧╩
%╟┴╠┴╦╘╔╦╔ ╙ ╥┴╙╙╘╧╤╬╔┼═ ╔ ┴╬┴╠╧╟╔▐╬╧┼ ╔┌═┼╬┼╬╔┼ ╙╦╧╥╧╙╘╔, ╙╫╤┌┴╬╬╧╩ ╙
%╒╟╠╧╫┘═ ╥┴╙╙╘╧╤╬╔┼═. Ё╧╫┼─┼╬╔┼ ╧┬┼╔╚ ╫┼╠╔▐╔╬ ╬┼═╧╬╧╘╧╬╬╧. э┴╦╙╔═╒═┘ ─╫╒╚
%╫┼╠╔▐╔╬ ═╧╟╒╘ ─╧╙╘╔╟┴╘╪╙╤ ╬┴ ╥┴┌╬┘╚ ╦╥┴╙╬┘╚ ╙═┼▌┼╬╔╤╚.
}
\label{sketch}
\end{figure}

Let the universe in our visualization (imagine it on the dome of a
planetarium) be filled (up to, say, $z\sim10$) with identical galaxies
of the same size. We see more distant galaxies as weak redden sources.
Angular size behaves according to Fig. \ref{graphs}: starting from some
distance farther sources look larger (they have smaller angular distance).
In dynamic, we would see that galaxies become redder and dimmer. However,
the main effect of recession would be visible due to diminishing angular
sizes of all galaxies. The rate of this diminishing of the angular size would
also go down (see Fig.  \ref{graphs} and a sketch in Fig. \ref{sketch}).
It is essential that according to an intuitive idea of a horizon, the dynamic
would ``freeze'' for the most distant sources. 

% Ё╒╙╘╪ ╫ ╬┴█┼╩ ╫╔┌╒┴╠╔┌┴├╔╔ (╨╥┼─╙╘┴╫╪╘┼ ┼┼ ╬┴ ╦╒╨╧╠┼ ╨╠┴╬┼╘┴╥╔╤)
%╫╙┼╠┼╬╬┴╤ ┌┴╨╧╠╬┼╬┴ (╫╨╠╧╘╪, ╙╦┴╓┼═, ─╧ $z\sim10$) ╔─┼╬╘╔▐╬┘═╔
%╟┴╠┴╦╘╔╦┴═╔ ╧─╬╧╟╧ ╨╧╙╘╧╤╬╬╧╟╧ ╥┴┌═┼╥┴. 
%т╧╠┼┼ ─┴╠┼╦╔┼ ╟┴╠┴╦╘╔╦╔ ═┘ ╫╔─╔═ ┬╧╠┼┼ ╙╠┴┬┘═╔ ╔ ╨╧╦╥┴╙╬┼╫█╔═╔.
%ї╟╠╧╫╧╩ ╥┴┌═┼╥ ╫┼─┼╘ ╙┼┬╤ ╙╧╟╠┴╙╬╧ ╥╔╙. \ref{graphs}: 
%╬┴▐╔╬┴╤ ╙ ╬┼╦╧╘╧╥╧╟╧ ╥┴╙╙╘╧╤╬╔╤ ┬╧╠┼┼ ─┴╠┼╦╔┼ 
%╟┴╠┴╦╘╔╦╔ ╫┘╟╠╤─╤╘ ┬\'╧╠╪█╔═╔ (╒╟╠╧╫╧┼ ╥┴╙╙╘╧╤╬╔┼ ─╧ ╬╔╚ ╨┴─┴┼╘). 
%ў ─╔╬┴═╔╦┼ ═┘ ┬╒─┼═ ╫╔─┼╘╪, ╦┴╦ ╟┴╠┴╦╘╔╦╔ ╙╘┴╬╧╫╤╘╙╤
%╫╙┼ ╦╥┴╙╬┼┼  ╔ ╙╠┴┬┼┼. я─╬┴╦╧ ╧╙╬╧╫╬╧╩ ▄╞╞┼╦╘ ``╥┴┌┬┼╟┴╬╔╤'' 
%┬╒─┼╘ ╙╧┌─┴╫┴╘╪╙╤ ╒═┼╬╪█┼╬╔┼═ ╒╟╠╧╫╧╟╧ ╥┴┌═┼╥┴
%╫╙┼╚ ╟┴╠┴╦╘╔╦, ╙╦╧╥╧╙╘╪ ╒═┼╬╪█┼╬╔╤ ╥┴┌═┼╥┴, ╬┴▐╔╬┴╤ ╙ ╬┼╦╧╘╧╥╧╟╧ ╥┴╙╙╘╧╤╬╔╤,
%╘┴╦╓┼ ┬╒─┼╘ ╨┴─┴╘╪ (╙═. ╟╥┴╞╔╦ ╬┴ ╥╔╙. \ref{graphs} ╔ ╦┴▐┼╙╘╫┼╬╬╒└ ╔╠╠└╙╘╥┴├╔└ ╬┴ ╥╔╙.
%\ref{sketch}). 
%є╒▌┼╙╘╫┼╬╬╧, ▐╘╧ ╫ ╙╧╧╘╫┼╘╙╘╫╔┼ ╙ ╔╬╘╒╔╘╔╫╬┘═ ╨╥┼─╙╘┴╫╠┼╬╔┼═ ╧ ╟╧╥╔┌╧╬╘┼ 
%╙╧┬┘╘╔╩  ─╔╬┴═╔╦┴ ┬╒─┼╘ ``┌┴═╔╥┴╘╪'' ─╠╤ ╙┴═┘╚ ─┴╠┼╦╔╚ (╔ ╙┴═┘╚ ╦╥┴╙╬┘╚) ╧┬▀┼╦╘╧╫. 

 As for the prospects of direct detection of dynamic of expansion, the first
results can be obtained due to $\dot z$ measurements with ultrastable
spectrographs on large telescopes of the next generation (see a review in 
\cite{quer}). Besides that, there are some hopes that GAIA can detect
decrease of sizes of gravitationally bound systems due to cosmic expansion
\cite{darling}. 

Expecting direct observational measurements of time
variations of quantities characterizing the  Hubble expansion, it is useful
to remind which parameters are observable directly and which, --- being
important for understanding and illustration of the expansion, --- represent
just a theoretical construction. This was the main goal of this note. 

%є ╘╧▐╦╔ ┌╥┼╬╔╤ ╥┼┴╠╪╬┘╚ ╬┴┬╠└─┼╬╔╩, ╫╔─╔═╧, ╨┼╥╫┘┼ ╥┼┌╒╠╪╘┴╘┘, 
%╦┴╙┴└▌╔┼╙╤ ╬┼╨╧╙╥┼─╙╘╫┼╬╬╧╩ ╥┼╟╔╙╘╥┴├╔╔
%─╔╬┴═╔╦╔ ╥┴╙█╔╥┼╬╔╤, ┬╒─╒╘ ╙╫╤┌┴╬┘ ╙ ╔┌═┼╥┼╬╔┼═ $\dot z$ ╙ ╨╧═╧▌╪└ ╒╠╪╘╥┴╙╘┴┬╔╠╪╬┘╚
%╙╨┼╦╘╥╧╟╥┴╞╧╫ ╬┴ ╘┼╠┼╙╦╧╨┴╚ ╬╧╫╧╟╧ ╨╧╦╧╠┼╬╔╤ (╙═. ╧┬┌╧╥, ╬┴╨╥╔═┼╥, ╫ \cite{quer}).
%%Quercellini et al.). 
%ы╥╧═┼ ╘╧╟╧, ┼╙╘╪ ╬┴─┼╓─┘, ▐╘╧ ╙╨╒╘╬╔╦ GAIA ╙═╧╓┼╘ ╧┬╬┴╥╒╓╔╘╪ ▄╞╞┼╦╘
% ╒═┼╬╪█┼╬╔╤ ╥┴┌═┼╥┴ ╟╥┴╫╔╘┴├╔╧╬╬╧-╙╫╤┌┴╬╬┘╚  ╙╔╙╘┼═ ┌┴ ╙▐┼╘ ╦╧╙═╧╠╧╟╔▐┼╙╦╧╟╧
%╥┴╙█╔╥┼╬╔╤ \cite{darling}. 
%ў ╨╥┼──╫┼╥╔╔ ╨╥╤═╧╟╧ ▄╦╙╨┼╥╔═┼╬╘┴╠╪╬╧╟╧ ╧┬╬┴╥╒╓┼╬╔╤ ╫╥┼═┼╬╬╧╩ ┌┴╫╔╙╔═╧╙╘╔ ╫┼╠╔▐╔╬, 
%╚┴╥┴╦╘┼╥╔┌╒└▌╔╚ ╚┴┬┬╠╧╫╙╦╧┼ ╥┴╙█╔╥┼╬╔┼, ╔═┼┼╘ ╙═┘╙╠ ┼▌┼ ╥┴┌ ╬┴╨╧═╬╔╘╪, ╦┴╦╔┼ ╫┼╠╔▐╔╬┘ ─╧╙╘╒╨╬┘
%╬┴┬╠└─┴╘┼╠└ ╬┼╨╧╙╥┼─╙╘╫┼╬╬╧, ┴ ╦┴╦╔┼, ╚╧╘╪ ╔ ╫┴╓╬┘ ─╠╤ ╨╧╬╔═┴╬╔╤ ╔ ┴─┼╦╫┴╘╬╧╟╧ ╧╨╔╙┴╬╔╤
%╥┴╙█╔╥┼╬╔╤ ╫╙┼╠┼╬╬╧╩, ╬╧ ╨╥┼─╙╘┴╫╠╤└╘ ╙╧┬╧╩ ▐╔╙╘╧ ╘┼╧╥┼╘╔▐┼╙╦╔╩ ╦╧╬╙╘╥╒╦╘. 
% ▄╘╧═ ╔ ╙╧╙╘╧╤╠┴ ├┼╠╪  ╬┴╙╘╧╤▌┼╩ ┌┴═┼╘╦╔.

In brief, our discussion can be summarized as follows:

%ы╥┴╘╦╧ ╬┴█┴ ┴╥╟╒═┼╬╘┴├╔╤ ╙╫╧─╔╘╙╤ ╦ ╙╠┼─╒└▌┼═╒:

\begin{itemize}
\item We want to define quantities which fit the best to our intuitive
understanding of visible distance and velocity of the Hubble flow in an
expanding universe.

  \item Proper distance is a fundamental quantity of the theory and does not
depend on our observational abilities and current astrophysical knowledge.  

  \item We see an object as it was at the moment of emission, so it is
natural to consider the distance at the moment of emission as an important
characteristic of the source. 

  \item Proper distance at the moment of emission can be calculated in the same
way as the angular distance. In addition, the angular distance and its
derivative correspond to our psychological perception of receding and to
intuitive expectations about the behavior of objects on the horizon. 
Therefore,  just angular distance and its derivative (in time) are the most
natural characteristics of the Hubble flow {\it from the observer's point of
view}.

\end{itemize}

%\begin{itemize}
%\item э┘ ╚╧╘╔═ ╧╨╥┼─┼╠╔╘╪ ╫┼╠╔▐╔╬┘, ═┴╦╙╔═┴╠╪╬╧ ╙╧╧╘╫┼╘╙╘╫╒└▌╔┼ ╔╬╘╒╔╘╔╫╬╧═╒ ╨╧╬╔═┴╬╔└
%╫╔─╔═┘╚ ╬┴┬╠└─┴╘┼╠┼═ ╥┴╙╙╘╧╤╬╔╤ ╔ ╙╦╧╥╧╙╘╔ ╚┴┬┬╠╧╫╙╦╧╟╧ ╨╧╘╧╦┴ ╫ ╥┴╙█╔╥╤└▌┼╩╙╤ ╫╙┼╠┼╬╬╧╩.

%  \item    є╧┬╙╘╫┼╬╬╧┼ ╥┴╙╙╘╧╤╬╔┼ ╤╫╠╤┼╘╙╤ ╫┘─┼╠┼╬╬┘═ ╫ ╘┼╧╥╔╔, ┴ ╘┴╦╓┼ ╬┼ ┌┴╫╔╙╤▌╔═ ╧╘ ╘┼╚╬╔▐┼╙╦╔╚ 
%╙╥┼─╙╘╫ ╬┴┬╠└─┼╬╔╩ ╔  ╬┴█╔╚ ╘┼╦╒▌╔╚  ┴╙╘╥╧╞╔┌╔▐┼╙╦╔╚ ╨╧┌╬┴╬╔╩.

% \item   Ї┴╦ ╦┴╦ ═┘ ╫╔─╔═ ╧┬▀┼╦╘ ╘┴╦╔═, ╦┴╦╔═ ╧╬ ┬┘╠, ╦╧╟─┴ ╔┌╠╒▐╔╠ ╨╥╔╬╔═┴┼═┘╩ ╬┴═╔ ╙╫┼╘,  
%┼╙╘┼╙╘╫┼╬╬╧ ╥┴╙╙═┴╘╥╔╫┴╘╪ ╥┴╙╙╘╧╤╬╔┼ 
%╬┴ ═╧═┼╬╘ ╔┌╠╒▐┼╬╔╤ (┼╟╧ ╬┴┬╠└─┴╘┼╠╪ ╫ ┬╧╠╪█┼╩ ╙╘┼╨┼╬╔ ``╫╔─╔╘'', ┴ ╬┼ ╫┘▐╔╙╠╤┼╘)
%╫ ╦┴▐┼╙╘╫┼ ╚┴╥┴╦╘┼╥╔╙╘╔╦╔ ╔╙╘╧▐╬╔╦┴.

%\item   є╧┬╙╘╫┼╬╬╧┼ ╥┴╙╙╘╧╤╬╔┼, ╔┌═┼╥┼╬╬╧┼ {\it ╫ ═╧═┼╬╘ ╔┌╠╒▐┼╬╔╤} 
%╨╥╔╬╔═┴┼═╧╟╧ ╙┼╩▐┴╙ ╙╔╟╬┴╠┴, ╫┘▐╔╙╠╤┼╘╙╤ ╨╧ ╘╧╩ ╓┼ ╞╧╥═╒╠┼,
%▐╘╧ ╔ ╒╟╠╧╫╧┼ ╥┴╙╙╘╧╤╬╔┼. ы╥╧═┼ ╘╧╟╧, ╙┴═╧ ╒╟╠╧╫╧┼ ╥┴╙╙╘╧╤╬╔┼ ╔ ╨╧╫┼─┼╬╔┼
%┼╟╧ ╨╥╧╔┌╫╧─╬╧╩ ╧╘╫┼▐┴└╘ ╦┴╦ ╨╙╔╚╧╠╧╟╔▐┼╙╦╧═╒ ╫╧╙╨╥╔╤╘╔└ ╒─┴╠┼╬╔└ ╧┬▀┼╦╘┴,
%╘┴╦ ╔ ╔╬╘╒╔╘╔╫╬╧ ╧╓╔─┴┼═╧═╒ ╨╧╫┼─┼╬╔└ ╧┬▀┼╦╘╧╫ ``╬┴ ╟╧╥╔┌╧╬╘┼''.
%є╧╧╘╫┼╘╙╘╫┼╬╬╧, ╔═┼╬╬╧ ╒╟╠╧╫╧┼ ╥┴╙╙╘╧╤╬╔┼ 
%╔ ┼╟╧ ╨╥╧╔┌╫╧─╬┴╤ ╨╧ ╫╥┼═┼╬╔ ╤╫╠╤└╘╙╤ ╬┴╔┬╧╠┼┼
%┼╙╘┼╙╘╫┼╬╬┘═╔ ╚┴╥┴╦╘┼╥╔╙╘╔╦┴═╔ ╚┴┬┬╠╧╫╙╦╧╟╧ ╨╧╘╧╦┴ {\it ╙ ╘╧▐╦╔ ┌╥┼╬╔╤ ╬┴┬╠└─┴╘┼╠╤}.

%\end{itemize} 

%я─╬┴╦╧ ╔┌═┼╬┼╬╔┼ ╒╟╠╧╫╧╟╧ ╥┴┌═┼╥┴ ╧┬▀┼╦╘╧╫ ╤╫╠╤┼╘╙╤ ╫┼╙╪═┴ ╔╠╠└╙╘╥┴╘╔╫╬┘═ ╨╥╔ ╔┌╠╧╓┼╬╔╔ ╧╙╬╧╫
%╦╧╙═╧╠╧╟╔╔ ╫ ╦┴▐┼╙╘╫┼ ╨╥╔═┼╥┴ ╨╧╘┼╬├╔┴╠╪╬╧ ╬┴┬╠└─┴┼═╧╟╧ ╙╫╧╩╙╘╫┴, ╧╘╥┴╓┴└▌┼╟╧ ╥┴╙█╔╥┼╬╔┼ ╫╙┼╠┼╬╬╧╩.
%ы ╙╧╓┴╠┼╬╔└, ╧┬┘▐╬╧ ╧╬╧ ╬┼─╧╙╘┴╘╧▐╬╧ ╧╘╥┴╓┼╬╧ ╫ ╒▐┼┬╬╧╩ ╠╔╘┼╥┴╘╒╥┼. 
%№╘╧╘ ╨╥╧┬┼╠ ═┘ ╔ ╨╧╙╘┴╥┴╠╔╙╪ ┌┴╨╧╠╬╔╘╪.

\vskip 0.1cm

\noindent
{\bf Acknowledgements}
We thank K.A. Postnov, S.A. Tyulbashev, V.A.Potapov and V.Sahni for discussions and comments. 
AT was supported by the RFBF grant 11-02-00643.
% э┘ ╨╥╔┌╬┴╘┼╠╪╬┘ ы.с. Ё╧╙╘╬╧╫╒ ╔ є.с. Ї└╠╪┬┴█┼╫╒ ┌┴ ╦╧══┼╬╘┴╥╔╔ ╔
%╧┬╙╒╓─┼╬╔╤. 
%Є┴┬╧╘┴ сЇ ┬┘╠┴ ╨╧──┼╥╓┴╬┴ ╟╥┴╬╘╧═ Єццщ 11-02-00643.

%\newpage

%\newpage

%\newpage

\end{document}